\documentclass[fleqn, preprint,review,12pt]{elsarticle}
\usepackage{lmodern}

\usepackage[margin=1in]{geometry}
\usepackage{physics}
\usepackage{amsmath}
\usepackage{amssymb}
\usepackage{esint}
\usepackage{siunitx}
\usepackage{multicol}
\usepackage{pdflscape}
\usepackage{multirow}
\usepackage{cleveref}
\usepackage{multicol}
\usepackage{booktabs}
\usepackage{float}
\usepackage{svg}
\usepackage{epstopdf}   
\usepackage{cuted}
\usepackage[draft]{changes}

\usepackage{makecell}

\usepackage{lineno}
\usepackage{caption}
\usepackage{subcaption}
\makeatletter
\let\LN@align\align
\let\LN@endalign\endalign
\renewcommand{\align}{\linenomath\LN@align}
\renewcommand{\endalign}{\LN@endalign\endlinenomath}
\let\LN@gather\gather
\let\LN@endgather\endgather
\renewcommand{\gather}{\linenomath\LN@gather}
\renewcommand{\endgather}{\LN@endgather\endlinenomath}
\makeatother

\newcolumntype{P}[1]{>{\centering\arraybackslash}p{#1}}
\newcolumntype{M}[1]{>{\centering\arraybackslash}m{#1}}

\newcommand{\tmean}[1]{\overline{#1}}
\newcommand{\enstmean}[1]{\left\langle \overline{#1} \right\rangle}
\usepackage{xcolor}
\usepackage[normalem]{ulem}

\newif\ifshowredst
\showredstfalse  

\usepackage{soul}
\sethlcolor{yellow}

\DeclareSIUnit\density{\kilogram\per\meter\tothe{3}}
\DeclareSIUnit\velocity{\meter\per\second}

\usepackage{soul}

\journal{}

\begin{document}
\begin{frontmatter}

\title{Reynolds-number-dependent entrance bias in temperature polarization of lab-scale membrane distillation}

\author[inst1]{Soyoung Kim}
\author[inst2]{Tiezheng Tong}
\cortext[cor1]{Corresponding author}
\author[inst1]{Yinuo Noah Yao\corref{cor1}}

\affiliation[inst1]{organization={Zachry Department of Civil and Environmental Engineering},
            addressline={Texas A\&M University}, 
            city={College Station},
            postcode={77843}, 
            state={TX},
            country={USA}}

\affiliation[inst2]{organization={School of Sustainable Engineering and the Built Environment},
            addressline={Arizona State University}, 
            city={Tempe},
            postcode={85281}, 
            state={AZ},
            country={USA}}

\begin{abstract}
Lab-scale membrane distillation (MD) modules
are commonly used for membrane characterization and
developing empirical heat-transfer correlations. Entrance-induced flow structures can introduce
geometry-specific effects into measured temperature
polarization (TP) and heat transfer. However, the dependence of these effects on module geometry and flow condition remains unclear. We performed three-dimensional computational fluid dynamics (CFD) simulations of direct contact membrane distillation (DCMD) over inlet Reynolds numbers of $\Re_{\mathrm{in}}=10$-$1000$, focusing on side-chamber length and main-channel height. We discovered that counter-rotating streamwise vortices formed near the main-channel entrance across all geometries, but their downstream persistence increased with $\Re_{\mathrm{in}}$. At lower $\Re_{\mathrm{in}}$, the vortices mostly decayed before mid-channel, while at higher $\Re_{\mathrm{in}}$ they persisted farther downstream and produced pronounced spatial variations in TP. Increasing side-chamber length reduced the secondary-flow kinetic energy entering the main channel, and generally strengthened TP, whereas increasing main-channel height reduced the bulk velocity and streamwise kinetic energy, resulting in a thicker thermal boundary layer and stronger TP. For most geometries, the deviation from the straight-channel baseline changed from stronger TP at lower $\Re_{\mathrm{in}}$ to weaker TP at higher $\Re_{\mathrm{in}}$. Entrance effects therefore vary in both magnitude and direction with operating condition and cannot be represented by a constant correction across the Reynolds-number range.
\end{abstract}

\end{frontmatter}


\section{Introduction}
\label{sec:intro}

Climate change and rising global populations have intensified water stress due to the growing imbalance between supply and demand~\cite{Hoekstra2014, Jones2019}. To address these challenges, membrane-based technologies have gained popularity due to their capability to recover water from unconventional water sources (e.g., seawater~\cite{Eykens2017, Kim2019}, brackish water~\cite{Lin2014, Pendergast2011}, and wastewater~\cite{Ali2024, Deshmukh2018, Kelly2025}). Recently, membrane distillation (MD) has been extensively studied in the context of desalination, particularly for removing salts from reverse osmosis brine, which is the waste stream that contains concentrated salts \cite{Lawson1997}. Since MD is thermally driven, low-grade heat sources (e.g., industrial waste heat, solar energy, etc.) can  be utilized to reduce energy consumption and the associated carbon footprint \cite{Ali2024, Khayet2011, Eykens2017, Kim2019, Chen2023}.

Accurate prediction of heat transfer and the associated temperature polarization (TP) is important for both lab-scale membrane characterization and full-scale MD modeling. TP occurs when the temperature difference across the membrane becomes smaller than that between the bulk feed and permeate streams~\cite{Bourawi2006, Diez1999, Anvari2020, Tijing2015, Gryta2008}, which leads to a reduction in the effective vapor-pressure driving force and, consequently, limits permeate flux. Additionally, membrane-surface temperatures also affect mineral scaling behavior, making accurate characterization of TP important for interpreting permeate-flux and scaling measurements in lab-scale MD experiments \cite{Christie2022-zd}. Given the challenges in measuring membrane surface temperatures at both lab-scale modules and full-scale systems, empirical Nusselt-Reynolds correlations are commonly used to characterize heat transfer and estimate TP in MD models \cite{Hitsov2015, Chen2009, Ali2013, Ali2024, warsinger2015, Dudchenko2020, Yao2023}. Reliable characterization of TP in lab-scale modules is therefore important for both evaluating performance of newly designed membranes and assessing the applicability of empirical heat-transfer relationships to larger systems.

However, hydrodynamics and heat transfer in lab-scale MD modules can be strongly affected by module geometry \cite{Yang2012, Shakaib2012, Kuang2019, Yao2023, Dudchenko2020}. Dudchenko \emph{et al.} \cite{Dudchenko2020} demonstrated that the design of lab-scale MD modules significantly affects the empirical correlations derived from experimental data. Yao \emph{et al.} \cite{Yao2023} subsequently attributed part of these discrepancies to entrance-induced hydrodynamics, which become increasingly important in smaller lab-scale modules. Fig. \ref{fig:DCMD_intro}(a) shows a typical lab-scale direct contact membrane distillation (DCMD) design, which consists of an inlet, a side chamber, a right-angled bend, and a main membrane channel. According to prior studies \cite{Yao2023, Lee2016, Salem2019}, such geometric configurations result in distinct hydrodynamic behaviors, such as flow separation, recirculation zones, and secondary vortices, which significantly affect local heat transfer and, consequently, TP on the membrane surface. For example, the combination of the inlet and side-chamber  resembles an impinging jet that enhances local heat transfer through mechanisms such as stagnation zones \cite{Liu1993, Webb1995}, secondary vortex formation \cite{Fox1993, Tesar2002, Song2007b}, and turbulence \cite{Gardon1965, Webb1995}. Similarly, the right-angled bends between the side chambers and main channel lead to the formation of Dean vortices \cite{Yao2023}. When these secondary flows persist in the main channel, they can increase transverse mixing and substantially alter local heat transfer and TP on the membrane surface \cite{Metzger1985, Dudchenko2020, Yao2023}.

For lab-scale membrane characterization, entrance-induced hydrodynamics can alter TP, influencing measured permeate flux and membrane-surface conditions relevant to scaling. Such variations complicate comparisons across studies, as measured differences in membrane performance may reflect both membrane properties and lab-scale hydrodynamics. For full-scale modeling, heat-transfer correlations derived from short lab-scale modules may contain geometry-specific contributions that are not representative of longer MD channels. Applying these correlations to larger systems may therefore introduce uncertainty in predicted performance. Although module-scale testing could reduce this uncertainty, testing each new membrane material at larger scales would require additional experimental effort and cost. Reducing geometry-specific entrance effects in lab-scale modules could help address both concerns. Yao \emph{et al.} \cite{Yao2023} showed that these geometries can generate Dean vortices in lab-scale MD modules and suggested suppressing these entrance-induced structures to reduce their influence on lab-scale correlations. However, the effects of individual geometric parameters on the strength and downstream persistence of these secondary flows were not systematically investigated. More importantly, it remains unclear whether suppressing secondary flows consistently reduces deviations from entrance-minimized channel behavior across different flow conditions.


In this study, three-dimensional computational fluid dynamics (CFD) simulations are performed to investigate the impacts of module geometry and Reynolds number on entrance-induced secondary flows and TP in lab-scale DCMD modules. We quantify the effects of individual geometric parameters on the strength and downstream persistence of these flows and compare the resulting TP with an ideal straight-channel baseline, which serves as a baseline with minimal entrance effects. The main study focuses on side-chamber length and main-channel height, which alter the flow through different mechanisms. Side-chamber height was studied as a sensitivity analysis, with the results provided in the Supplementary Information (Fig.~\ref{fig:appendix_TPC_HS}). By varying these parameters over a range of Reynolds numbers, we assess whether the geometry that minimizes the deviation from straight-channel behavior changes with flow conditions. The results identify conditions under which lab-scale temperature polarization is more representative of the heat-transfer behavior of entrance-minimized baseline. \Cref{sec:goven_eqs} presents the governing equations and numerical methods, while \Cref{sec:geo-parmeters,sec:setup} describe the geometric configurations and simulation setup. \Cref{sec:results} presents the effects of geometry and Reynolds number on TP and secondary-flow behavior, followed by a discussion of the underlying mechanisms and implications for lab-scale DCMD design in \Cref{sec:discuss1}.

\section{Methodology}
\label{sec:method}
\subsection{Governing equations}
\label{sec:goven_eqs}

\begin{figure}
\centerline{\includegraphics[scale=0.5]{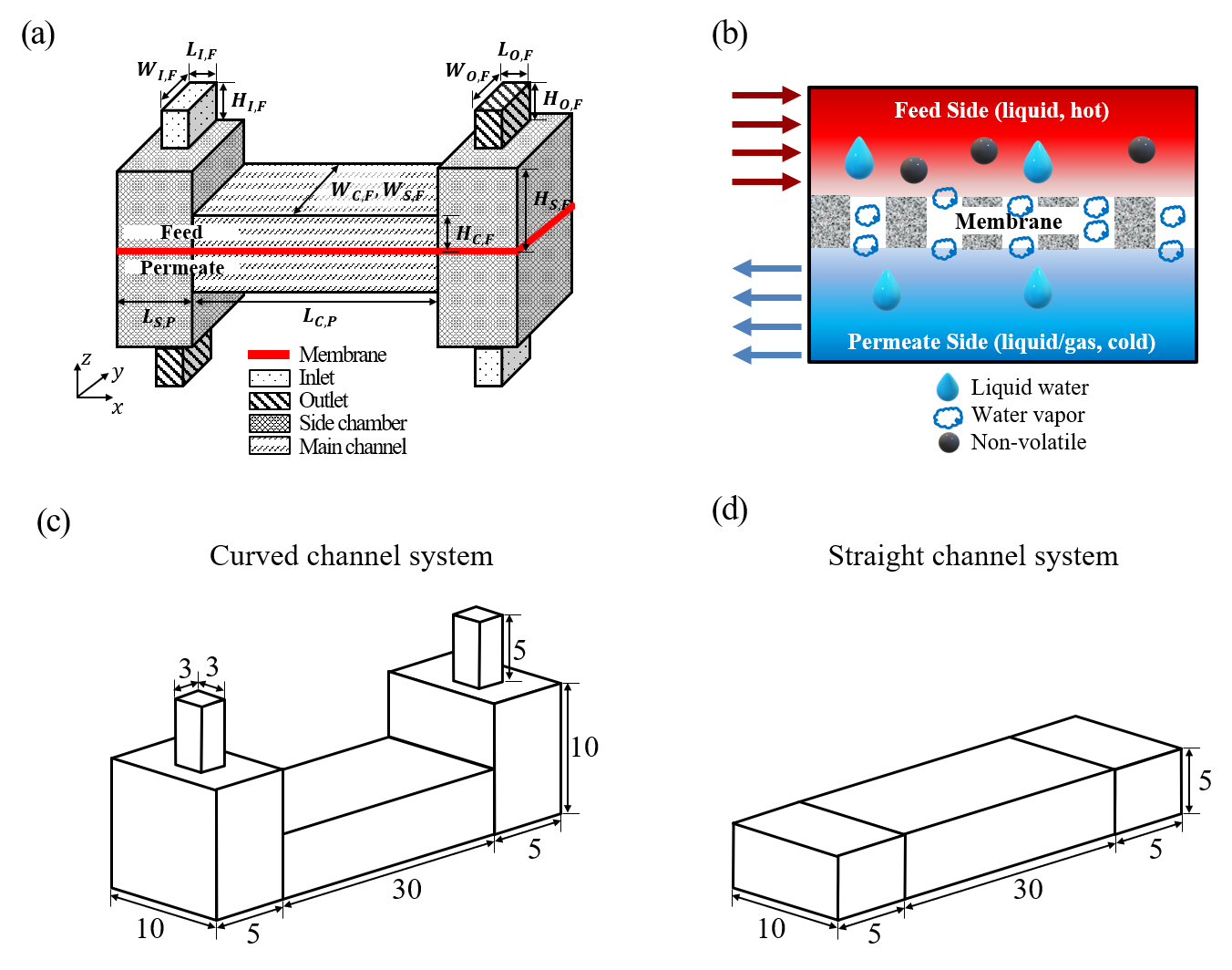}}
\caption{Schematic showing (a) the lab-scale DCMD configuration considered in this study, (b) the DCMD process, (c) the reference lab-scale module geometry, and (d) the ideal straight-channel geometry used as the entrance-minimized baseline. All dimensions in (c) and (d) are in mm.}
\label{fig:DCMD_intro}
\end{figure}

We consider membrane distillation of feedwater containing sodium chloride. In a typical counter-current DCMD configuration, a hot stream flows in the feed side, while a cold stream moves in the opposite direction on the permeate side (Fig.~\ref{fig:DCMD_intro}(b)). The temperature difference between the two streams drives water vapor through the pores of the hydrophobic membrane to the permeate side, while non-volatile solutes such as sodium chloride are retained on the feed side. 

Hydrodynamics, heat transport, and mass transport are simulated in both the feed and permeate domains, with the subscript $j=\{F,P\}$ denoting the respective sides. Following Yao \emph{et al.} \cite{Yao2023}, we solve the compressible Navier-Stokes equations to account for variations in solution density and viscosity with respect to temperature and salt concentration. The equations are given by

\begin{subequations}
\label{eq:whole1}
\begin{equation}
        \pdv{\rho_j}{t} + \divergence{\left(\rho_j\mathbf{u}_j\right)} =0 \\ 
\end{equation}

\begin{equation}
        \pdv{\left(\rho_j \mathbf{u}_j\right)}{t} + \divergence{\left(\rho_j\mathbf{u}_j\bf{u}_j\right)} = -\grad{p}_j + \divergence{\mathbf{\tau}_j} + \rho_j\mathbf{g} \\
\end{equation}

\begin{equation}
        \mathbf{\tau}_j = \mu_j \left( \grad\mathbf{u}_j + \left(\grad\mathbf{u}_j\right)^{\textrm{T}} - \frac{2}{3}\left(\divergence{\mathbf{u}_j}\right)\mathbf{I}\right),
\end{equation}
\end{subequations}

\noindent where $\rho_j$~[\unit{\density}] is the solution density, $\mathbf{u}_j$~[\unit{\velocity}]$ = \begin{pmatrix} u_j & v_j & w_j \end{pmatrix}^T$ is the velocity vector, $p_j$~[\unit{\pascal}] is the total pressure, $g$~[\unit{\meter\per\second\tothe{2}}]$ = \begin{pmatrix} 0 & 0 & -9.81 \end{pmatrix}$ is the gravitational acceleration vector, $\mathbf{\tau}_j$ is the viscous stress tensor, $\mu_j$~[\unit{\kilogram\per\meter\per\second}] is the solution dynamic viscosity, and $\mathbf{I}$~is the identity matrix.

The temperature in the MD module is computed from the energy equations

\begin{subequations}
\label{eq:whole2}
\begin{equation}
        \pdv{\left(\rho_j h_j\right)}{t} + \divergence{\left(\rho_j \mathbf{u}_j h_j\right)} = \divergence{\left(k_j\grad T_j\right)} + \pdv{p_j}{t} + \mathbf{u}_j \cdot \grad{p_j}\\ 
\end{equation}

\begin{equation}
        h_j = \int_{273.15\unit{\kelvin}}^{T_j} c_{p,j}\left(T,C_j\right)\,\mathrm{d}T,
\end{equation}
\end{subequations}

\noindent where $k_j$~[\unit{\watt\per\meter\per\kelvin}] is the thermal conductivity,  $T_j$~[\unit{\kelvin}] is the temperature, $h_j$~[\unit{\joule\per\kilogram}] is the specific enthalpy, $c_{p,j}$~[\unit{\joule\per\kilogram\per\kelvin}] is the specific heat capacity, and $C_j$~[\si{\kilogram\per\meter\cubed}] is the sodium chloride concentration. The thermophysical properties of the solution at different salt concentrations and temperatures are evaluated as described in Yao \emph{et al.} \cite{Yao2023}.

The transport of sodium chloride is governed by the advection-diffusion equation 

\begin{equation}
    \label{eq:scalar_eq}
    \pdv{C_j}{t} + \divergence{\left(\mathbf{u}_j C_j \right)} = \divergence{\left(D_j \grad{C_j}\right)},
\end{equation}

\noindent where $D_j$~[\unit{\meter\tothe{2}\per\second}] is the molecular diffusion coefficient.

The computational boundaries are represented by $\Gamma_{i,j}$, where $i=\{I,O,W,M\}$ refers to the inlet, outlet, wall, and membrane boundaries, respectively, and $j=\{F,P\}$ refers to the feed and permeate sides. For example, $\Gamma_{I,F}$ denotes the feed-side inlet. At the membrane, the normal velocity is determined by the transmembrane permeate flux $J_w$ [\unit{\kilogram\per\meter\tothe{2}\per\second}] as

\begin{subequations}
\label{eq:whole3}
\begin{equation}
        \mathbf{u}_F(\mathbf{x}, t) = \frac{J_w}{\rho_F} \mathbf{n}_F, \quad \mathbf{x} \in \Gamma_{M,F} \label{subeq:flux1} \\ 
\end{equation}

\begin{equation}
        \mathbf{u}_P(\mathbf{x}, t) = -\frac{J_w}{\rho_P} \mathbf{n}_P, \quad \mathbf{x} \in \Gamma_{M,P}, \label{subeq:flux2}\\
\end{equation}
\end{subequations}

\noindent where $\mathbf{n}_F$ and $\mathbf{n}_P$ are the outward unit normal vectors of the feed and permeate fluid domains at the membrane, respectively. The flux $J_w$, which is driven by the transmembrane vapor pressure difference due to the temperature gradient across the membrane, is modeled as a linear function of the vapor pressure difference such that

\begin{equation}
    \label{eq:flux_Jw}
    J_w = A_w\left(p_{M,F} - p_{M,P}\right),
\end{equation}

\noindent where $A_w$~[\unit{\second\per\meter}] is the vapor permeability, which depends on temperature, membrane surface pressure, and intrinsic membrane properties~\citep{Bouchrit2015}. The partial vapor pressure on each side of the membrane, $p_{M,j}$~[\unit{\pascal}], is calculated as 

\begin{subequations}
\label{eq:whole4}
\begin{equation}
        p_{M,j} \left(C_{M,j}, T_{M,j}\right) = a_{w,j} \left(C_{M,j}\right) p_{sat,j} \left(T_{M,j}\right) \\ 
\end{equation}

\begin{equation}
        a_{w,j} = 1 - 0.03122m + 0.001482m^2 \\
\end{equation}

\begin{equation}
        p_{sat,j} = \exp\left(23.238 - \frac{3841}{T_{M,j} - 45}\right),\\
\end{equation}
\end{subequations}

\noindent where $a_{w,j}$~[-] is the water activity, expressed as a function of solute concentration, $m$~[\unit{\mole\per\kilogram}] is the sodium chloride molality, $T_{M,j}$~[\unit{\kelvin}] and $C_{M,j}$~[\unit{\kg\per\meter\tothe{3}}] are the membrane surface temperature and concentration on side $j$, respectively, and $p_{sat,j}$~[\unit{\pascal}] is the saturation vapor pressure, which depends on temperature and is estimated using the Antoine equation \cite{zhang2022, Smith1950}. In this formulation, temperature and concentration polarization are considered through the local membrane-surface temperatures and concentrations used to calculate the vapor-pressure difference.

At the membrane, sodium chloride transport is governed by a balance between diffusive and convective fluxes \cite{Lou2019, Lou2021, Yao2023, Ling2019, Ling2021} such that

\begin{subequations}
\label{eq:whole5}
\begin{equation}
        \mathbf{n}_F \cdot D_F\grad\mathbf{C}_F(\mathbf{x}, t)  = - \dfrac{J_w}{\rho_F} C_{M,F}, \quad \mathbf{x} \in \Gamma_{M,F} \label{subeq:bc1} \\ 
\end{equation}

\begin{equation}
        \mathbf{n}_P \cdot D_P\grad\mathbf{C}_P(\mathbf{x}, t)  =  \dfrac{J_w}{\rho_P} C_{M,P}, \quad \mathbf{x} \in \Gamma_{M,P}, \label{subeq:bc2} \\ 
\end{equation}
\end{subequations}

\noindent where $C_{M,F}$~[\si{\kilogram\per\meter\cubed}] and $C_{M,P}$~[\si{\kilogram\per\meter\cubed}] are the sodium chloride concentrations at the membrane surface on the feed and permeate sides, respectively.

For temperature, Neumann boundary conditions are imposed to account for both convective and conductive heat transport \citep{Sarti1985, Lou2019, Lou2021, Yao2023} as

\begin{subequations}
\label{eq:whole6}
\begin{equation}
        \mathbf{n}_F \cdot k_F\grad\mathbf{T}_F(\mathbf{x}, t)= -\lambda J_w - h_M \left(T_{M,F} - T_{M,P}\right), \quad \mathbf{x} \in \Gamma_{M,F} \label{subeq:n_bc1} \\ 
\end{equation}

\begin{equation}
        \mathbf{n}_P \cdot k_P\grad\mathbf{T}_P(\mathbf{x}, t) = \lambda J_w + h_M \left(T_{M,F} - T_{M,P}\right), \quad \mathbf{x} \in \Gamma_{M,P}, \label{subeq:n_bc2} \\ 
\end{equation}
\end{subequations}

\noindent where $\lambda$~[\unit{\joule\per\kilogram}] is the latent heat of water,  and $h_M$~[\unit{\watt\per\meter\tothe{2}\per\kelvin}] is the effective membrane thermal conductance. For pressure, zero-gradient Neumann boundary conditions are applied.

At the feed and permeate inlets, velocity, temperature, and concentration are prescribed as $\mathbf{u}_{I,j}$, $T_{I,j}$, and $C_{I,j}$, respectively, while a zero-gradient condition is applied to pressure. At the outlets, zero-gradient conditions are imposed for velocity, temperature, and concentration, and pressure is set to zero. Along the solid walls, no-slip conditions are applied to velocity, while zero-gradient conditions are imposed for pressure, temperature, and concentration. All boundary conditions are summarized in Table~\ref{tab:MD_BC}. The numerical method is verified against the simulations of Lou \emph{et al.} \cite{Lou2019} and validated against the experimental data of Bush \emph{et al.} \cite{Bush2016}. Details are provided in \Cref{sec:validation}.

\begin{landscape}
\begin{table}[H]
    \centering
    \caption{Summary of the boundary conditions applied in simulations of membrane distillation systems.}
    \begin{tabular}{c|c|c|c|c|c}
    \multicolumn{2}{c|}{}  & Velocity & Pressure & Concentration & Temperature\\ \hline
    \multirow{4}{*}{Feed} & Walls & $\mathbf{u}_F = 0$ & $ \mathbf{n}_F \cdot \grad{p_F} = 0$ & $\mathbf{n}_F \cdot \grad{C_F}  = 0 $ & $\mathbf{n}_F \cdot \grad{T_F} = 0$ \\
     & Inlet & $\mathbf{u}_F = \mathbf{u}_{I,F}$ & $ \mathbf{n}_F \cdot \grad{p_F} = 0$ & $C_F  = C_{I,F} $ & $T_F = T_{I,F}$ \\
     & Outlet & $\mathbf{n}_F \cdot  \grad{\mathbf{u}_F} = 0$ & $p_F = 0$ & $\mathbf{n}_F \cdot \grad{C_F}  = 0 $ & $\mathbf{n}_F \cdot \grad{T_F} = 0$ \\
     & Membrane & $\mathbf{u}_F = \dfrac{J_w}{\rho_F} \mathbf{n}_F$ & $ \mathbf{n}_F \cdot \grad{p_F} = 0$ & $\mathbf{n}_F \cdot D_F\grad\mathbf{C}_F = - \dfrac{J_w}{\rho_F} C_{M,F}$ & $\mathbf{n}_F \cdot k_F\grad\mathbf{T}_F = -\lambda J_w - h_M \left(T_{M,F} - T_{M,P}\right)$\\ \hline
    \multirow{4}{*}{Permeate} & Walls & $\mathbf{u}_P = 0$ & $ \mathbf{n}_P \cdot \grad{p_P} = 0$ & $\mathbf{n}_P \cdot \grad{C_P}  = 0 $ & $\mathbf{n}_P \cdot \grad{T_P} = 0$ \\
     & Inlet & $\mathbf{u}_P = \mathbf{u}_{I,P}$ & $ \mathbf{n}_P \cdot \grad{p_P} = 0$ & $C_P  = C_{I,P} $ & $T_P = T_{I,P}$ \\
     & Outlet & $\mathbf{n}_P \cdot  \grad{\mathbf{u}_P} = 0$ & $p_P = 0$ & $\mathbf{n}_P \cdot \grad{C_P}  = 0 $ & $\mathbf{n}_P \cdot \grad{T_P} = 0$ \\
     & Membrane & $\mathbf{u}_P = -\dfrac{J_w}{\rho_P} \mathbf{n}_P$ & $ \mathbf{n}_P \cdot \grad{p_P} = 0$ & $\mathbf{n}_P \cdot D_P\grad\mathbf{C}_P = \dfrac{J_w}{\rho_P} C_{M,P}$ & $\mathbf{n}_P \cdot k_P\grad\mathbf{T}_P = \lambda J_w + h_M \left(T_{M,F} - T_{M,P}\right)$\\ \hline
    \end{tabular}    
    \label{tab:MD_BC}
\end{table}
\end{landscape}

\subsection{\label{sec:geo-parmeters}Geometric configurations}

In this study, we investigate lab-scale DCMD modules consisting of an inlet, an outlet, a side chamber, and a main channel (Fig.~\ref{fig:DCMD_intro}(a)). Each component is characterized by its length $L_{k,j}$, height $H_{k,j}$, and width $W_{k,j}$, where $k=\{I,O,S,C\}$ refers to the inlet, outlet, side chamber, and main channel, respectively, and $j=\{F,P\}$ refers to the feed and permeate sides.

We first consider a reference configuration  (Fig.~\ref{fig:DCMD_intro}(c)) previously shown to generate significant unsteady Dean vortices \cite{Yao2023}. The inlet ($L_{I,j}\times W_{I,j}\times H_{I,j}$) and outlet dimensions ($L_{O,j}\times W_{O,j}\times H_{O,j}$) are $3\times3\times5$~\unit{\mm}, the side chamber dimensions ($L_{S,j}\times W_{S,j}\times H_{S,j}$) are $5\times10\times10$~\unit{\mm}, and the main channel dimensions ($L_{C,j}\times W_{C,j}\times H_{C,j}$) are $30\times10\times5$~\unit{\mm}. Two geometric parameters are varied relative to this reference configuration in the main analysis, the side-chamber length ($L_{S,j}$) and the main-channel height ($H_{C,j}$). 

Given the prohibitive computational cost of resolving thermal boundary layers in full-scale MD systems, we use an ideal straight channel (lacking any entrance-related geometries) as an entrance-minimized baseline to quantify geometry-induced deviations. For the reference geometry, the straight channel geometry ($L_{C,j}\times W_{C,j}\times H_{C,j}$) has dimensions $40\times10\times5$~\unit{\mm} and the same cross-section as the reference main channel, but contains no side chambers, inlets, outlets, or right-angled bends (Fig.~\ref{fig:DCMD_intro}(d)). For cases with varying $H_{C,j}$, the comparison with this reference straight-channel baseline includes changes in both entrance-induced flow and intrinsic main-channel transport.

\begin{figure}
\centerline{\includegraphics[scale=0.4]{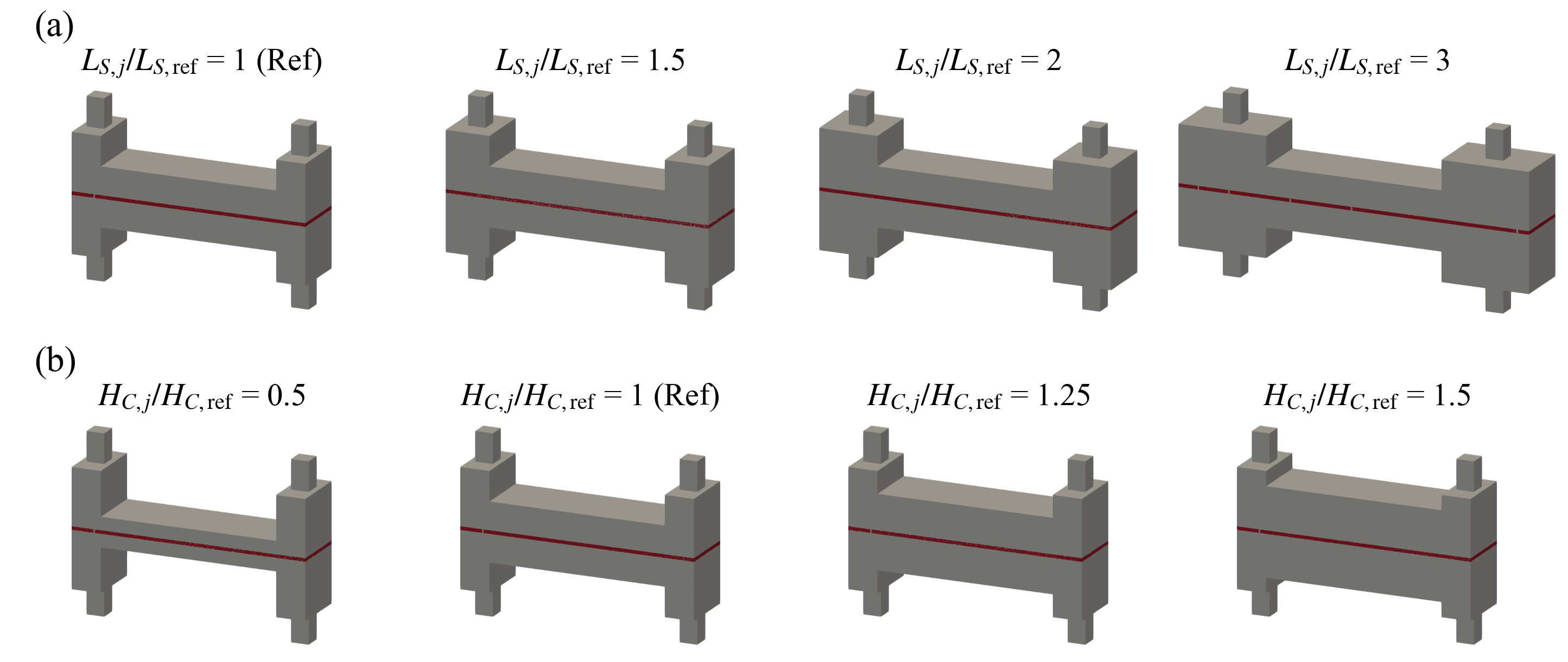}}
\caption{\label{fig:DCMD_arr_dim}Schematic illustrating variations in (a) side-chamber length ($L_{S,j}/L_{S,\mathrm{ref}}$) and (b) main-channel height ($H_{C,j}/H_{C,\mathrm{ref}}$). The red surface indicates the membrane location. All other geometric dimensions are fixed at their reference values.}
\end{figure}

To assess the impacts of module geometry, the side-chamber length was varied as $L_{S,j}/L_{S,\mathrm{ref}}=\{1,1.5,2,3\}$, with $L_{S,\mathrm{ref}}=5$~\unit{\mm} (Fig.~\ref{fig:DCMD_arr_dim}(a)). We also varied the main-channel height as $H_{C,j}/H_{C,\mathrm{ref}}=\{0.5,1,1.25,1.5\}$, with $H_{C,\mathrm{ref}}=5$~\unit{\mm} (Fig.~\ref{fig:DCMD_arr_dim}(b)). The active membrane area is $4~\si{cm^2}$ for all cases except those with varying side-chamber length, where it ranges from $4$ to $6~\si{cm^2}$. The side-chamber height was additionally studied using $H_{S,j}/H_{S,\mathrm{ref}}=\{1,2,3\}$, with $H_{S,\mathrm{ref}}=10$~\unit{\mm}. Since changes in $H_{S,j}$ only resulted in  minor differences in temperature polarization over the range considered, these results are provided in the Supplementary Information (Fig.~\ref{fig:appendix_TPC_HS}). 

\subsection{\label{sec:setup}Simulation setup}

The inlet temperatures on the feed and permeate sides were set to $T_{I,F} = 313.15~\unit{\kelvin}$ and $T_{I,P} = 293.15~\unit{\kelvin}$, respectively, which resulted in a temperature difference of 20~\unit{\kelvin}. The vapor permeability was set to $A_w = \num{2.96e-06}$~\unit{\second\per\meter}, following the two-dimensional calibration against experimental permeate-flux measurements reported \cite{Lou2019,Yao2023,Bush2016}. The membrane transport parameters were constant across all simulations. The effective membrane thermal conductance was set to $h_M=825.7$~\si{\watt\per\meter\squared\per\kelvin}, and both membrane transport parameters were held constant across all simulations. The sodium chloride concentration at the feed inlet ($C_{I,F}$) was 35~\unit{\kilogram\per\meter\tothe{3}}, and 0~\unit{\kilogram\per\meter\tothe{3}} at the permeate inlet ($C_{I,P}$). The inlet velocity at the feed side was defined as $\mathbf{u}_{I,F} = (0, 0, -u_{in})$, while on the permeate side it was $\mathbf{u}_{I,P} = (0, 0, u_{in})$. 

In this study, we define inlet Reynolds numbers as

\begin{equation}
    \Re_{\mathrm{in}} = \frac{u_{in} D_{h,\text{in}}}{\nu_0},
\end{equation}

\noindent where $u_{in}$ is the inlet velocity, $D_{h,\text{in}}$~[\unit{\meter}] is the hydraulic diameter of the inlet, and $\nu_0 = \num{1e-6}$~\unit{\meter\tothe{2}\per\second} is the kinematic viscosity of water. We performed simulations at $\Re_{\mathrm{in}}=\{10,50,100,250,500,1000\}$, with $u_{in}=0.333$~\unit{\meter\per\second} at $\Re_{\mathrm{in}}=1000$.

To determine the grid resolution required for accurate simulations, a grid-convergence study was performed using the reference configuration (\ref{sec:convergence}). The grid was uniformly spaced in the $x$-, $y$-, and $z$-directions with $\Delta x = \Delta y = \Delta z_{\max} = \num{1.25e-4}$~\unit{\meter}, with additional refinement applied near the membrane. An adaptive time-step scheme was used to ensure numerical stability, with the maximum Courant number defined as

\begin{equation}
    CFL_{\max} = \max \left( \frac{u \Delta t}{\Delta x} + \frac{v \Delta t}{\Delta y} + \frac{w \Delta t}{\Delta z} \right).
\end{equation}

\noindent The time step was adjusted to maintain $CFL_{\max} \leq 0.8$, where $\Delta t$~[\unit{\second}] is the time-step size. A characteristic convective timescale was defined based on the main channel as

\begin{equation}
\tau = \frac{L_{C,F} \mathcal{A}_c}{u_{\mathrm{in}}\mathcal{A}_{\mathrm{in}}},
\end{equation}

\noindent where $\mathcal{A}_{in}$~[\unit{\meter\tothe{2}}] is the inlet area, and $\mathcal{A}_c$~[\unit{\meter\tothe{2}}] is the cross-sectional area of the main channel, respectively. The simulations were continued until the time-spatial-averaged temperature polarization coefficient reached a statistically stationary state. The required physical simulation time varied with $\Re_{\mathrm{in}}$ because the main-channel convective timescale increases as the flow rate decreases. 




\subsection{\label{sec:TPC} Quantification of temperature polarization and hydrodynamics}

Most hydrodynamic analyses presented below are reported for the feed side. The permeate stream flows counter-currently, with the streamwise direction opposite to that of the feed side, and is not discussed separately unless otherwise noted.

Two comparisons are used to quantify the effects of module geometry. First, we compare each modified geometry with the reference module to isolate the effect of the individual geometric parameter. Second, the resulting behavior is compared with an ideal straight channel, which serves as a baseline with minimal entrance effects, to quantify the entrance-induced deviation from straight-channel behavior. All comparisons are performed at the same inlet flow rate.

To quantify TP, the dimensionless temperature polarization coefficient ($\mathbb{TPC}$) is defined as the ratio of the effective membrane temperature gradient to the overall temperature difference in the bulk fluids. The equation is given by

\begin{equation}
    \label{eq:tpc}
    \mathbb{TPC}(x,y,t) = \frac{T_{M,F}(x,y,t) - T_{M,P}(x,y,t)}{T_{I,F} - T_{I,P}}.
\end{equation}

\noindent Based on the definition in Eq.~(\ref{eq:tpc}), $\mathbb{TPC}= 1$ indicates the absence of temperature polarization, while smaller $\mathbb{TPC}$ indicates stronger polarization. To compare the spatial evolution of flow and transport across geometries with different dimensions, we define a normalized feed-side coordinate $\xi$ relative to the chamber-to-channel junction $x_{\mathrm{J},F}$ as

\begin{equation}
\xi =
\begin{cases}
\dfrac{x-x_{\mathrm{J},F}}{L_{S,F}},
& x_{\mathrm{J},F}-L_{S,F} \le x < x_{\mathrm{J},F}
\quad \text{(side chamber)}, \\[6pt]
\dfrac{x-x_{\mathrm{J},F}}{L_{C,F}},
& x_{\mathrm{J},F} \le x \le x_{\mathrm{J},F}+L_{C,F}
\quad \text{(main channel)}.
\end{cases}
\label{eq:xi}
\end{equation}

\noindent Thus, $-1\le\xi<0$ corresponds to the feed-side side chamber, $\xi=0$ refers to the chamber-to-channel junction, and $0\le\xi\le1$ corresponds to the feed-side main channel.

To characterize the temporal and spatial variations in temperature polarization, we define the time-averaged temperature polarization coefficient ($\tmean{\mathbb{TPC}}$), the time-spatial-averaged temperature polarization coefficient ($\langle\overline{\mathbb{TPC}}\rangle$), and the time-spanwise-averaged temperature polarization coefficient ($\langle\overline{\mathbb{TPC}}\rangle_y$) as

\begin{subequations}
\label{eq:whole8}

\begin{equation}
\overline{\mathbb{TPC}}(x,y)
=\frac{1}{N_t}\sum_{n=1}^{N_t}\mathbb{TPC}(x,y,t_n),
\label{subeq:TPC1}
\end{equation}

\begin{equation}
\langle\overline{\mathbb{TPC}}\rangle
=\frac{1}{N_xN_y}\sum_{l=1}^{N_x}\sum_{m=1}^{N_y}\overline{\mathbb{TPC}}(x_l,y_m),
\label{subeq:TPC2}
\end{equation}

\begin{equation}
\langle\overline{\mathbb{TPC}}\rangle_y(x)
=\frac{1}{N_y}\sum_{m=1}^{N_y}\overline{\mathbb{TPC}}(x,y_m),
\label{subeq:TPC3}
\end{equation}
\end{subequations}

\noindent where $N_t$ is the number of time snapshots used for time averaging, $(x_l,y_m)$ are the spatial coordinates, and $t_n$ is the sampling time. The indices are $l=\{1,2,\dots,N_x\}$ and $m=\{1,2,\dots,N_y\}$ with $N_x$ and $N_y$ representing the numbers of points in the $x$ and $y$ directions, respectively. Time-averaged quantities were calculated over the statistically stationary portion of each simulation using snapshots sampled at a uniform physical-time interval. At least 15 snapshots were included in the averaging window for each case.

To quantify the entrance-induced secondary flow, we define the normalized instantaneous streamwise vorticity as
\begin{equation}
\omega_x^\dagger(x_c,y,z,t)=\frac{D_{h,\mathrm{in}}}{u_{\mathrm{in}}}\omega_x(x_c,y,z,t),
\qquad
\omega_x=(\nabla\times\mathbf{u})_x,
\end{equation}



\noindent where $\omega_x=(\nabla\times\mathbf{u})_x$ is the streamwise vorticity and $x_c$ is the streamwise location of the cross-section of interest.  The normalized time-averaged streamwise vorticity used to visualize the coherent vortex structures is

\begin{equation}
\overline{\omega_x^\dagger}(x_c,y,z)=\frac{1}{N_t}\sum_{n=1}^{N_t}\omega_x^\dagger(x_c,y,z,t_n).
\end{equation}

\noindent The time-averaged cross-sectional streamwise-vorticity magnitude is defined as

\begin{equation}
\overline{\langle |\omega_x^{\dagger}| \rangle_{\mathcal{A}}}=\frac{1}{N_t}\sum_{n=1}^{N_t}\frac{1}{\mathcal{A}}\int_\mathcal{A}\left|\omega_x^\dagger(x_c,y,z,t_n)\right|\,d\mathcal{A},
\end{equation}



\noindent where $\mathcal{A}$ is the local cross-sectional area. The relative strengths of the streamwise and secondary flows were further quantified using time-spatial-averaged kinetic energies such that

\begin{equation}
\left\langle \overline{k_{\mathrm{str}}}\right\rangle(x)
=\frac{1}{\mathcal{A} N_t}\sum_{n=1}^{N_t}\int_\mathcal{A} \frac{1}{2}u^2\,\mathrm{d}\mathcal{A},
\qquad
\left\langle \overline{k_{\mathrm{sec}}}\right\rangle(x)
=\frac{1}{\mathcal{A} N_t}\sum_{n=1}^{N_t}\int_\mathcal{A} \frac{1}{2}\left(v^2+w^2\right)\,\mathrm{d}\mathcal{A}.
\label{eq:k}
\end{equation}

\noindent where $\left\langle \overline{k_{\mathrm{str}}}\right\rangle$ and $\left\langle \overline{k_{\mathrm{sec}}}\right\rangle$ are the time-spatial-averaged streamwise and secondary-flow specific kinetic energies, respectively. 

\section{\label{sec:results}Results}

\subsection{\label{sec:all_cases} Effects of module geometry and Reynolds number on average temperature polarization} 

\begin{figure}
\centerline{\includegraphics[scale=0.3]{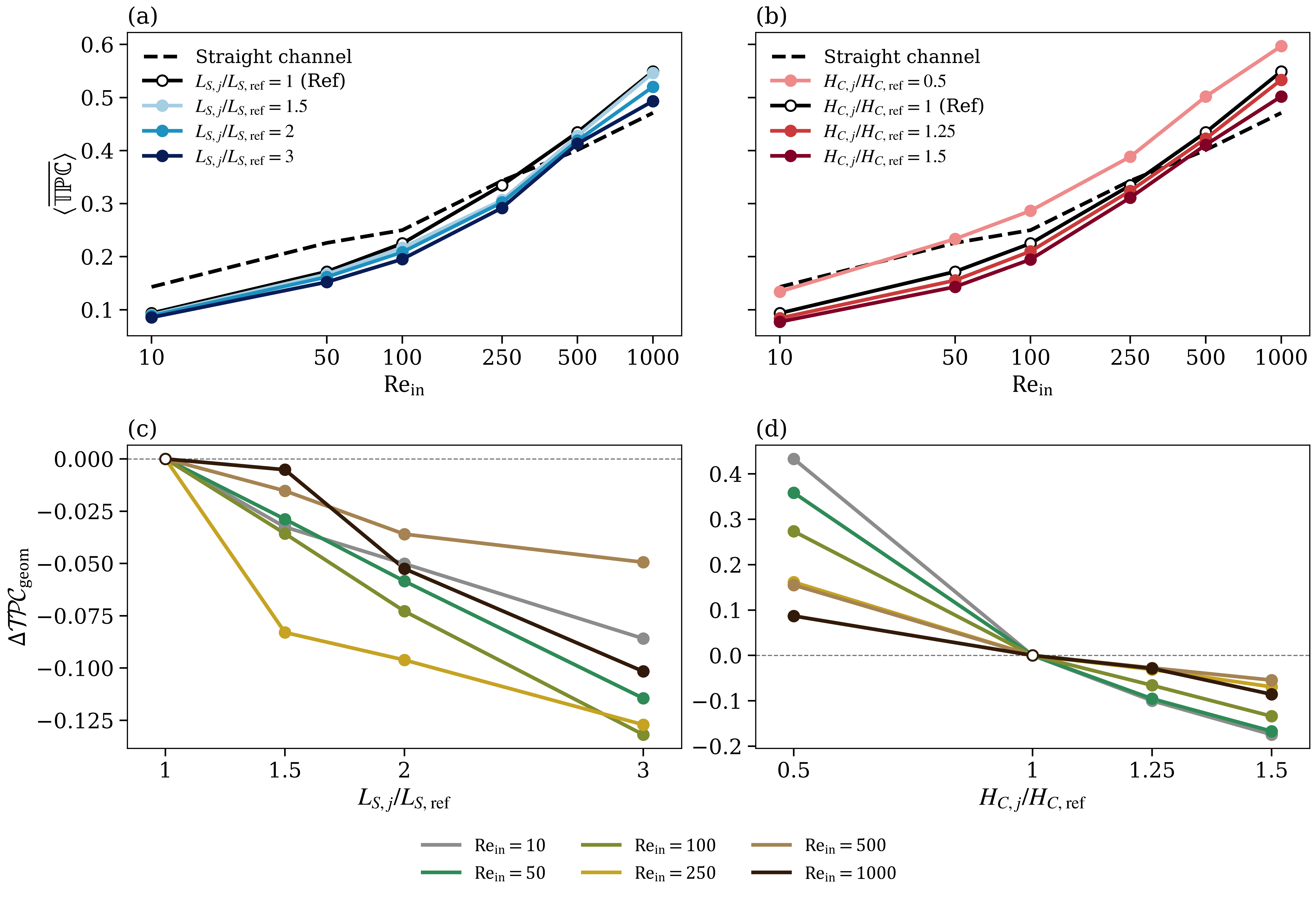}}
\caption{\label{fig:TPC_all}
Time-spatial-averaged temperature polarization coefficient ( $\enstmean{\mathbb{TPC}}$) as a function of $\Re_{\mathrm{in}}$ for varying (a) side-chamber length and (b) main-channel height. The dashed line refers to the ideal straight-channel baseline. Panels (c) and (d) show the corresponding relative difference from the reference module ($\Delta\mathcal{TPC}_{\mathrm{geom}}$) for varying side-chamber length and main-channel height, respectively.}
\end{figure}

In this section, we quantify the effects of the geometric parameters on the temperature polarization coefficient. Fig.~\ref{fig:TPC_all} shows the average temperature polarization coefficient ($\enstmean{\mathbb{TPC}}$) as a function of the inlet Reynolds number ($\Re_{\mathrm{in}}$) for both geometric parameters and the straight-channel baseline with minimal entrance effects (dotted line). For all cases, $\enstmean{\mathbb{TPC}}$ increases monotonically with $\Re_{\mathrm{in}}$, indicating a reduction in temperature polarization. The straight-channel baseline exhibits the same overall trend, showing that the increase in $\enstmean{\mathbb{TPC}}$ with $\Re_{\mathrm{in}}$ is not unique to the entrance geometry.

The impact of the entrance geometry is evident from the deviation of the lab-scale modules from the straight-channel baseline. Across the geometries considered (Fig.~\ref{fig:TPC_all}(a) for side-chamber length and Fig.~\ref{fig:TPC_all}(b) for main-channel height), a similar Reynolds-number-dependent trend is observed. At lower $\Re_{\mathrm{in}}$, the modules show lower $\enstmean{\mathbb{TPC}}$ than the straight-channel baseline, indicating stronger TP. At higher $\Re_{\mathrm{in}}$, the modules show higher $\enstmean{\mathbb{TPC}}$ (weaker TP) than the straight-channel baseline, with the deviation increasing with $\Re_{\mathrm{in}}$.  All modules except $H_{C,j}/H_{C,\mathrm{ref}} = 0.5$ shift from below to above the straight-channel baseline as $\Re_{\mathrm{in}}$ increases. Thus, the deviation from the straight-channel case changes in both magnitude and direction with $\Re_{\mathrm{in}}$. This shows that the entrance-induced deviation is not monotonic and can occur in either direction (improving or reducing temperature polarization) depending on the flow conditions.  

To quantify the effects of each geometric parameter on $\enstmean{\mathbb{TPC}}$, we compute the relative difference $\Delta{\mathcal{TPC}}_{\mathrm{geom}}$ at each $\Re_{\mathrm{in}}$ as

\begin{equation}
\Delta{\mathcal{TPC}}_{\mathrm{geom}} = \frac{\enstmean{\mathbb{TPC}}-\enstmean{\mathbb{TPC}}_{\mathrm{ref}}}{\enstmean{\mathbb{TPC}}_{\mathrm{ref}}}.
\end{equation}

\noindent Fig.~\ref{fig:TPC_all}(c) and (d) show the relative difference as a function of $L_{S,j}/L_{S,\mathrm{ref}}$ and $H_{C,j}/H_{C,\mathrm{ref}}$ across Reynolds numbers, respectively. Increasing $L_{S,j}/L_{S,\mathrm{ref}}$ results in a decrease in $\enstmean{\mathbb{TPC}}$ and therefore strengthens temperature polarization for all $\Re_{\mathrm{in}}$ except $\Re_{\mathrm{in}}=10$. At $\Re_{\mathrm{in}}=10$, the geometry-induced differences are negligible. For the main-channel height, $\enstmean{\mathbb{TPC}}$ generally decreases with increasing $H_{C,j}/H_{C,\mathrm{ref}}$, while the shallowest channel ($H_{C,j}/H_{C,\mathrm{ref}}=0.5$) results in a higher $\enstmean{\mathbb{TPC}}$ than the reference case across the Reynolds-number range considered.

\subsection{\label{sec:Space_TPC} Spatial variations in temperature polarization}

\begin{figure}
\centerline{\includegraphics[scale=0.5]{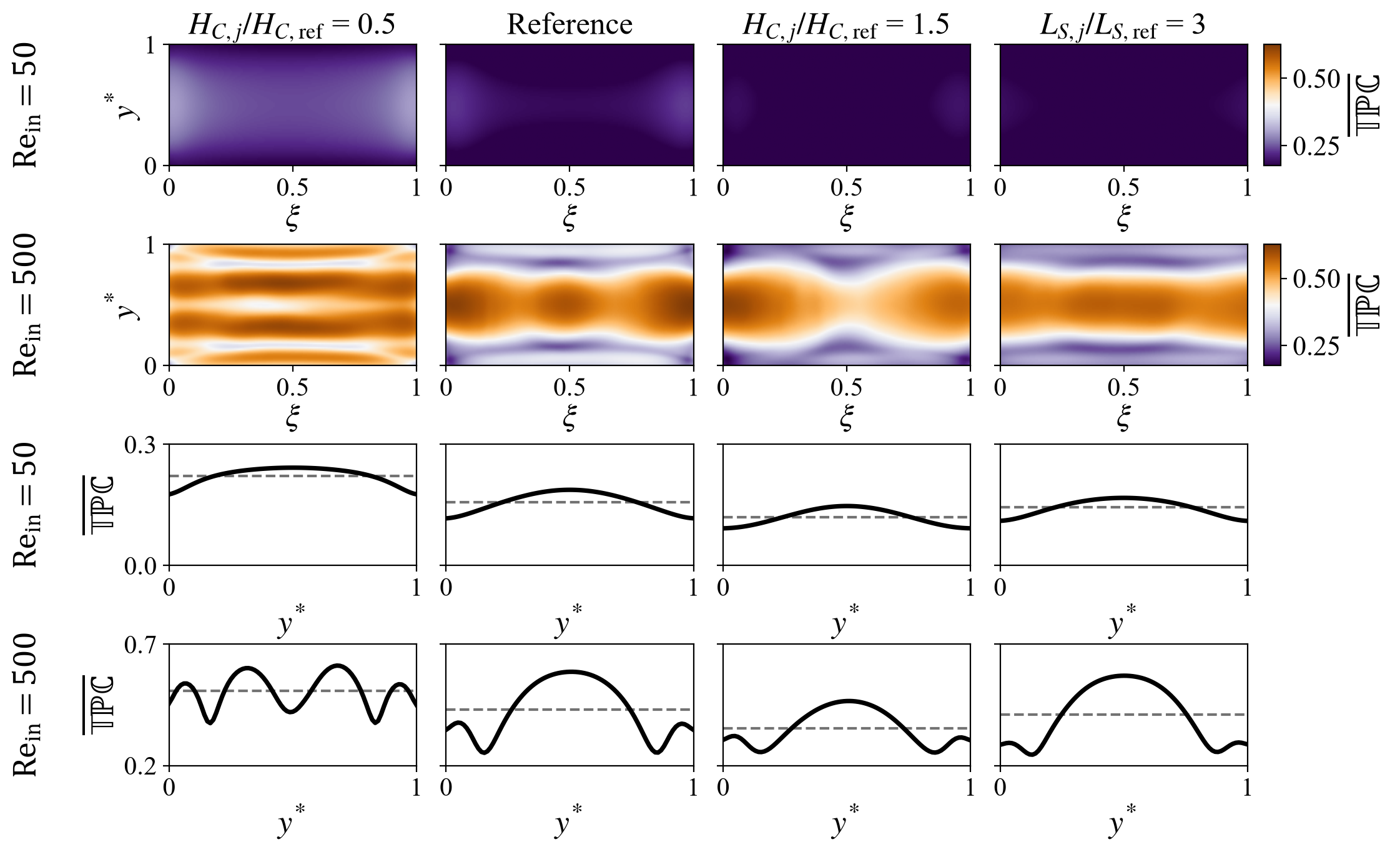}}
\caption{\label{fig:Case_TPC} 
Time-averaged temperature polarization coefficient ($\tmean{\mathbb{TPC}}$) for four representative geometries at $\Re_{\mathrm{in}}=50$ and 500. The first and second rows show membrane-surface distributions at $\Re_{\mathrm{in}}=50$ and 500, respectively. The third and fourth rows show the corresponding spanwise profiles at mid-channel ($\xi=0.5$).}
\end{figure}

To evaluate the impact of module geometry and flow conditions on the spatial distribution of temperature polarization, the time-averaged temperature polarization coefficient ($\tmean{\mathbb{TPC}}$) was computed for all simulated geometries and Reynolds numbers. Fig.~\ref{fig:Case_TPC} shows the membrane-surface contours of $\tmean{\mathbb{TPC}}$ and the corresponding spanwise profiles at mid-channel ($\xi=0.5$) for four representative geometries at $\Re_{\mathrm{in}}=50$ and 500. The membrane-surface contours and spanwise profiles for all simulated cases are provided in Supplementary Information Figs.~\ref{fig:appendix_TPC_mem} and~\ref{fig:appendix_TPC_center}  

At $\Re_{\mathrm{in}} = 50$, all simulations show small variations in $\tmean{\mathbb{TPC}}$ along the spanwise direction (Fig.~\ref{fig:Case_TPC}). The spanwise profiles at mid channel show the same behavior, with only small deviations from the spanwise-averaged value. Thus, at lower Reynolds numbers, modification in geometries primarily  modifies the overall magnitude of $\enstmean{\mathbb{TPC}}$, while having a negligible effect on its spanwise distribution.

At $\Re_{\mathrm{in}}=500$, significant spanwise variations are observed across the membrane surface (Fig.~\ref{fig:Case_TPC}). In particular, reducing the main-channel height to $H_{C,j}/H_{C,\text{ref}} = 0.5$ results in a distinct multi-peak structure, whereas the other three cases retain a dominant maximum near mid-span. This trend is consistent with the spanwise profiles at mid channel, with multiple peaks for $H_{C,j}/H_{C,\text{ref}} = 0.5$ and one dominant peak for the remaining cases. These results show that module geometry affects not only the magnitude of $\enstmean{\mathbb{TPC}}$ but also its spatial distribution at higher Reynolds numbers.

\subsection{\label{sec:flow} Secondary-flow persistence and spatial temperature polarization}

In \Cref{sec:Space_TPC}, we show weak spanwise variations in $\tmean{\mathbb{TPC}}$ at lower Reynolds numbers and significant spatial variations and structures at higher Reynolds numbers. To understand whether this difference is associated with the downstream persistence of the entrance-induced Dean vortices, we compare the normalized, time-averaged streamwise vorticity $\omega_x^{\dagger}$ at $\xi=0.05$ and $0.5$.

At the upstream location ($\xi=0.05$), counter-rotating streamwise Dean vortices are present for the representative geometries at both $\Re_{\mathrm{in}}=50$ and 500 (Fig.~\ref{fig:Vor1}). Vorticity contours for all simulated geometries and Reynolds numbers are provided in the Supplementary Information (Fig. \ref{fig:appendix_TPC_vort}). These Dean vortices are generated as the flow turns through the right-angled bend and enters the main channel \cite{Yao2023}. Their formation at both Reynolds numbers indicates that vortex formation alone cannot explain the different temperature polarization patterns observed on the membrane.

\begin{figure}
\centerline{\includegraphics[scale=0.3]{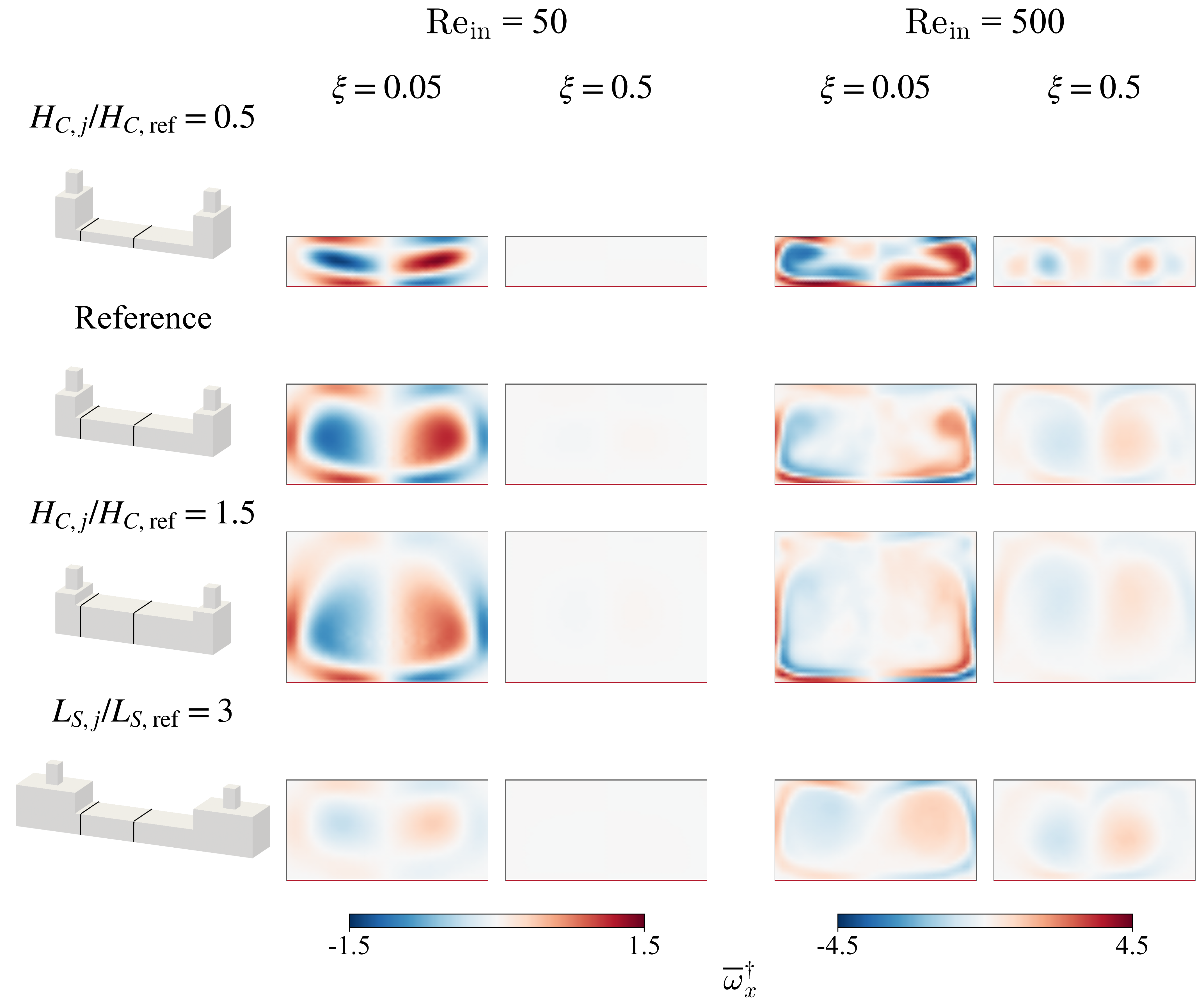}}
\caption{\label{fig:Vor1} Normalized time-averaged streamwise vorticity ($\omega_x^{\dagger}$) at $\xi=0.05$ (upstream) and $0.5$ (mid-channel) for four representative geometries at $\Re_{\mathrm{in}}=50$ and 500.}
\end{figure}

\begin{figure}
\centerline{\includegraphics[scale=0.35]{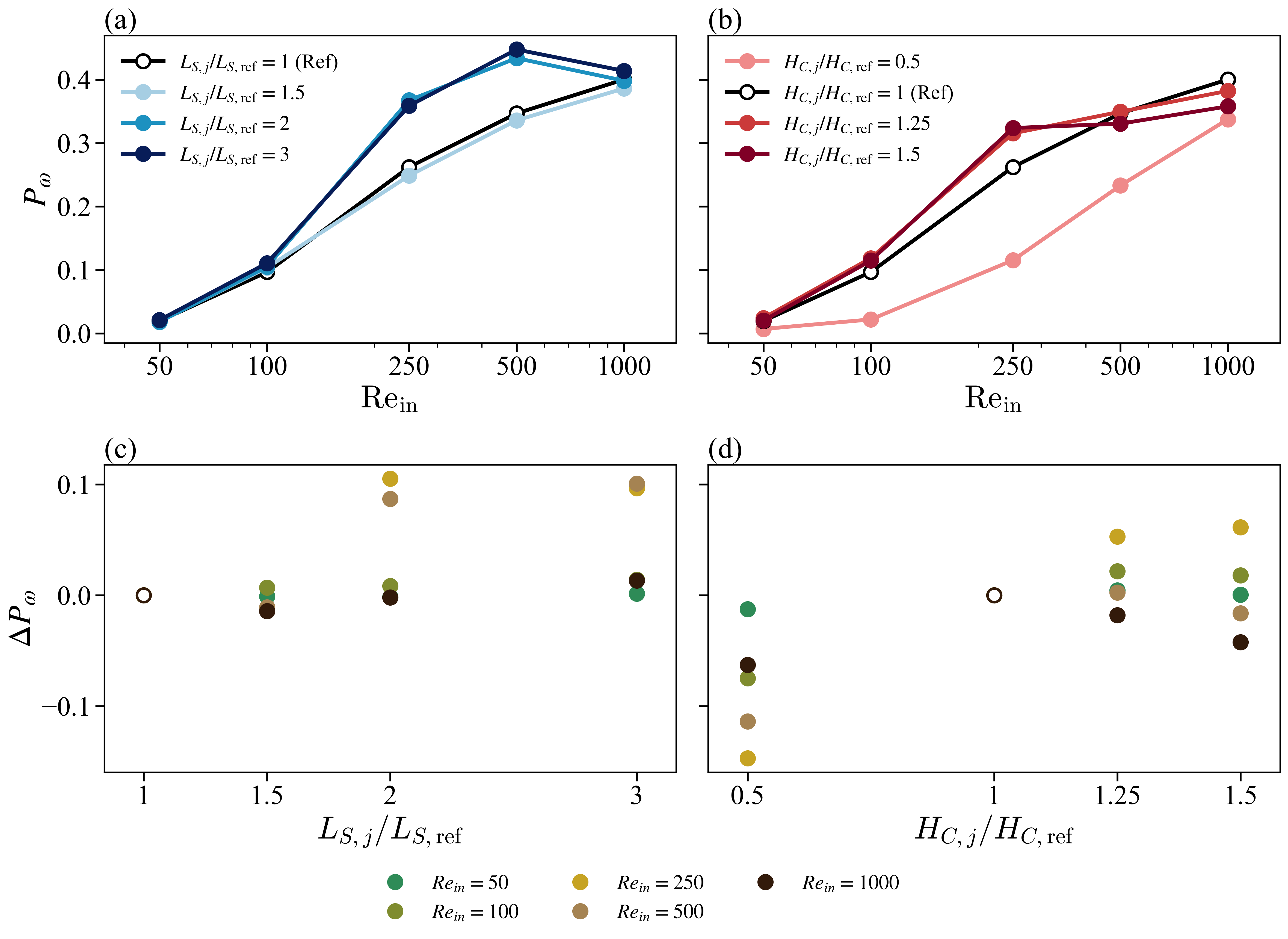}}
\caption{\label{fig:Vor2} Vorticity persistence ratio ($P_{\omega}$) as a function of $\Re_{\mathrm{in}}$ for varying (a) side-chamber length and (b) main-channel height, and the
corresponding difference with respect to the reference module ( $\Delta P_{\omega}$) for varying (c) side-chamber length and (d) main-channel height.}
\end{figure}

The downstream persistence of these vortices, however, differs substantially between the two Reynolds numbers. At $\Re_{\mathrm{in}}=50$, the Dean vortices have mostly decayed by mid-channel ($\xi=0.5$) for all representative geometries. At $\Re_{\mathrm{in}}=500$, counter-rotating Dean vortices remain in the middle of the channel (Fig.~\ref{fig:Vor1}). We further quantify the downstream persistence using the vorticity persistence ratio $P_{\omega}$ as

\begin{equation}
P_{\omega}=\frac{\left.\overline{\langle |\omega_x^{\dagger}| \rangle_\mathcal{A}}\right|_{\xi=0.5}}{\left.\overline{\langle |\omega_x^{\dagger}| \rangle_\mathcal{A}}\right|_{\xi=0.05}}.
\label{eq:persistence}
\end{equation}

\noindent Here, the magnitude of the normalized streamwise vorticity is first averaged over the cross section and then over time. As shown in Fig.~\ref{fig:Vor2}(a) and (b), $P_{\omega}$ increases with $\Re_{\mathrm{in}}$ for all geometries, indicating greater downstream persistence of the streamwise-vorticity magnitude at higher Reynolds numbers. To investigate the effect of geometry on persistence, we computed the deviation from the reference case as $\Delta P_{\omega} = P_{\omega} - P_{\omega,\mathrm{ref}}$. Fig.~\ref{fig:Vor2}(c) and (d) show that the effects of geometry depend on the flow condition. Increasing $L_{S,j}$ generally results in higher $\Delta P_{\omega}$, particularly at intermediate Reynolds numbers, although the dependence is not monotonic across all cases. For $H_{C,j}$, the shallowest channel ($H_{C,j}/H_{C,\mathrm{ref}}=0.5$) consistently shows lower persistence, while increasing $H_{C,j}$ above the reference generally increases $\Delta P_{\omega}$ at lower and intermediate Reynolds numbers. This trend becomes less consistent at $\Re_{\mathrm{in}}=1000$, which may be associated with greater unsteadiness of the vortex structures at higher Reynolds numbers \cite{Yao2023}.

These differences are consistent with the trends of temperature polarization on the membrane. At $\Re_{\mathrm{in}}=50$, the vortices have decayed by mid-channel, resulting in only weak spanwise variations in $\tmean{\mathbb{TPC}}$. At $\Re_{\mathrm{in}}=500$, the vortices remain, and significant spanwise variations are observed in  $\tmean{\mathbb{TPC}}$. In particular, $H_{C,j}/H_{C,\text{ref}}=0.5$ consists of multiple streamwise vortex structures at mid-channel and exhibits the multi-peak $\tmean{\mathbb{TPC}}$ distribution shown in Fig.~\ref{fig:Case_TPC}. In the other geometries, a dominant counter-rotating pair remains, and the corresponding $\tmean{\mathbb{TPC}}$ profile retains a single dominant maximum near mid-span. Instead of merely the formation of Dean vortices, these results show that both the downstream persistence and structure of the vortices are associated with the spatial variation of temperature polarization on the membrane.

\subsection{\label{sec:energy} Effects of geometry on flow and temperature polarization}

Sections~\ref{sec:all_cases}--\ref{sec:flow} show that geometric parameters and the Reynolds number affect both the magnitude and spatial distribution of temperature polarization. In particular, the downstream persistence of the secondary flow has a significant impact on the spatial variations of temperature polarization. In this section, we investigate how the streamwise and secondary-flow kinetic energies vary with geometry and the Reynolds number, and relate these changes to temperature polarization.

\begin{figure}
\centerline{\includegraphics[scale=0.25]{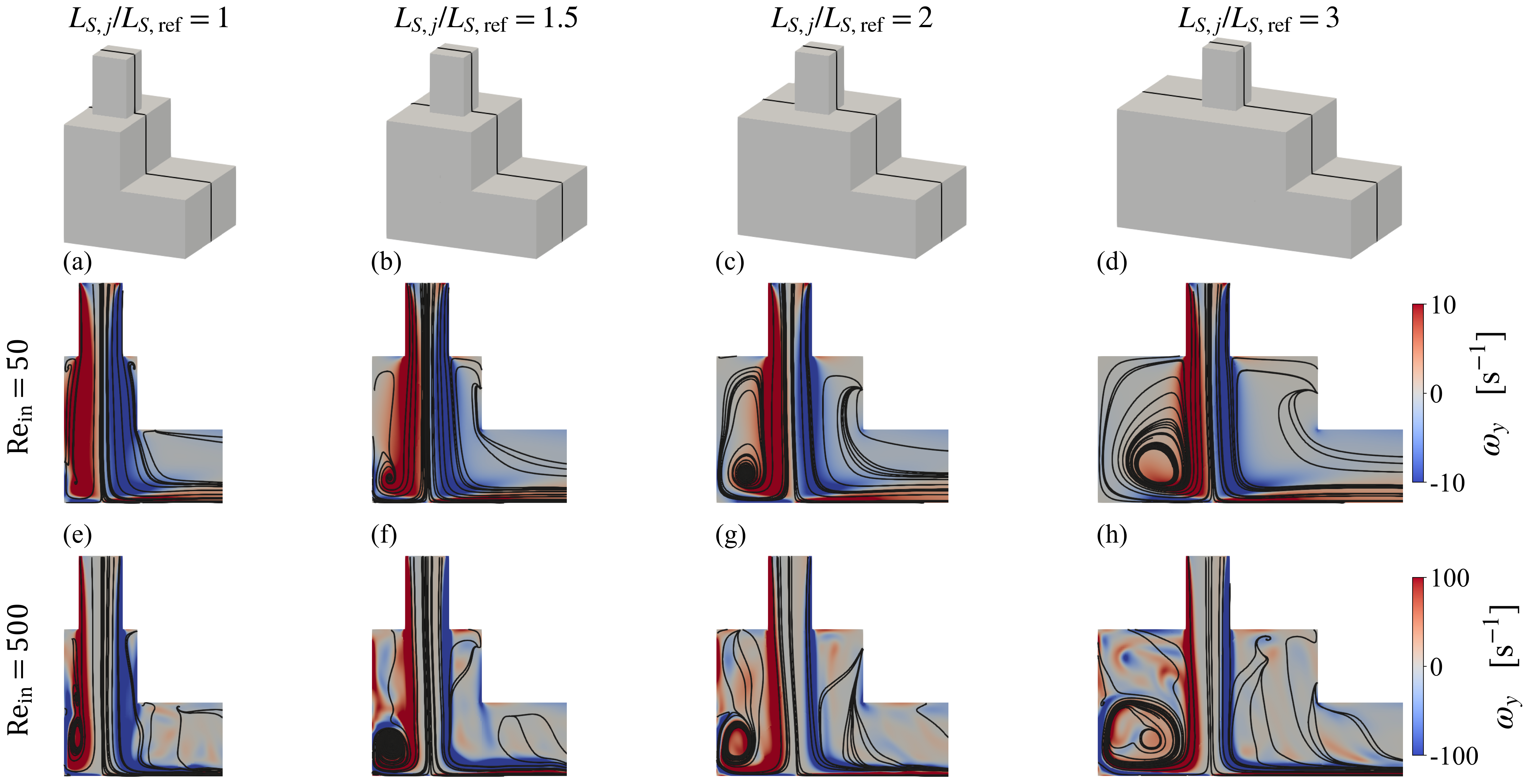}}
\caption{\label{fig:Case1_stream}Instantaneous streamlines and \(y\)-direction vorticity, \(\omega_y\), for (a, e) \(L_{S,j}/L_{S,\mathrm{ref}}=1\), (b, f) 1.5, (c, g) 2, and (d, h) 3, with the respective geometries shown in the top row. The middle and bottom rows correspond to \(\Re_{\mathrm{in}}=50\) and \(500\), respectively.}
\end{figure}

Fig.~\ref{fig:Case1_stream} shows the $y$-direction vorticity overlay with instantaneous streamlines for varying side-chamber lengths. As $L_{S,j}/L_{S,\mathrm{ref}}$ increases, we observed the formation of additional recirculation in the side chamber before the flow reaches the bend and the main channel. Similar recirculating structures have been reported for confined impinging-jet flows \cite{Kumar2025, Shuja2003-jf}. The additional recirculation increases flow interaction with the chamber walls and provides a possible pathway for greater viscous dissipation before the flow enters the main channel \cite{Shuja2003-jf}.

\begin{figure}
\centerline{\includegraphics[scale=0.3]{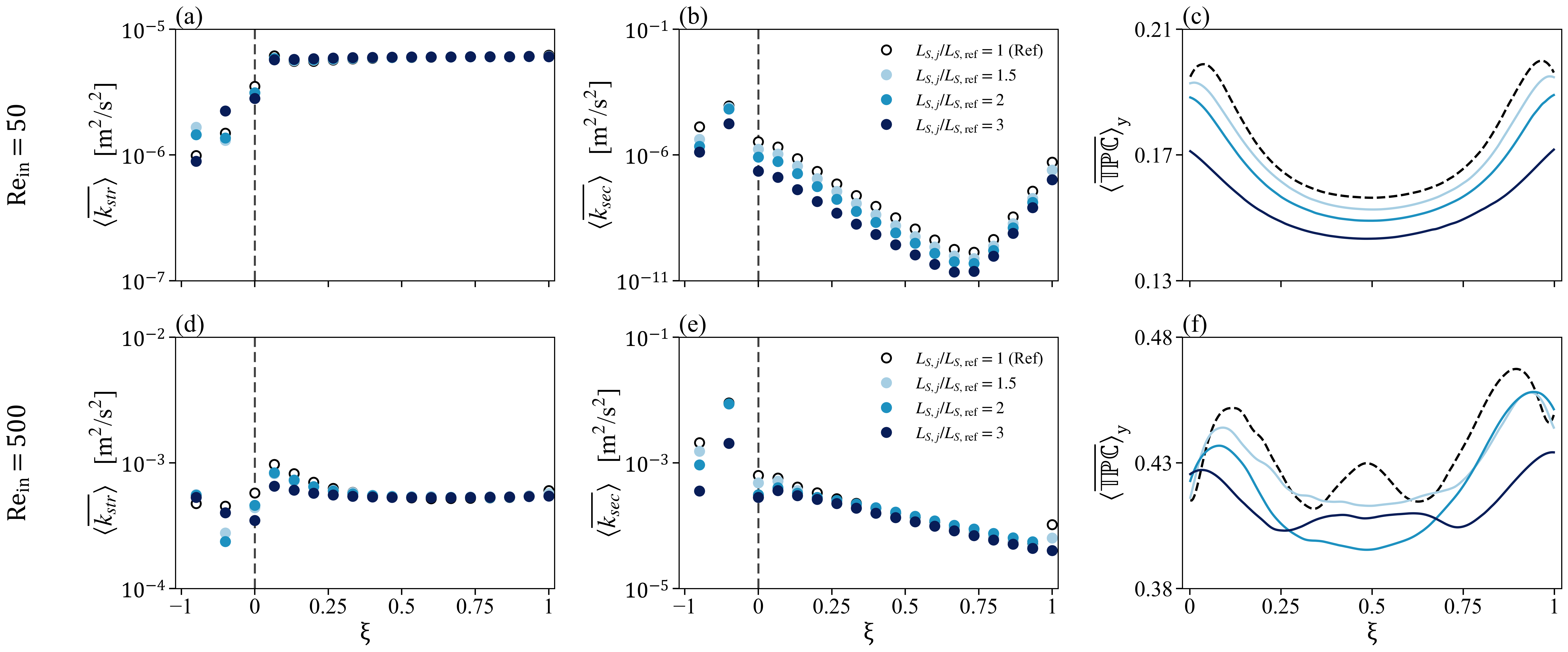}}
\caption{\label{fig:Energy1} Streamwise and secondary-flow kinetic energies and temperature polarization for varying side-chamber length at $\Re_{\mathrm{in}}=50$ (top) and 500 (bottom), respectively. (a, d) time-spatial-averaged streamwise kinetic energy ($\langle \overline{k_{\mathrm{str}}}\rangle$)
(b, e) time-spatial-averaged secondary-flow kinetic energy, ($\langle \overline{k_{\mathrm{sec}}}\rangle$), and
(c, f) time-spanwise-averaged temperature polarization coefficient ($\langle \overline{\mathbb{TPC}}\rangle_y$
). Negative $\xi$ corresponds to the side chamber, and the dashed line denotes the chamber-to-channel junction at $\xi=0$.}
\end{figure}

To quantify the corresponding changes, we computed the time-averaged streamwise and secondary-flow kinetic energies $\langle \overline{k_{\mathrm{str}}} \rangle$ and $\langle \overline{k_{\mathrm{sec}}} \rangle$
, and compared them with the time-spanwise-averaged temperature polarization coefficient, $\langle \overline{\mathbb{TPC}} \rangle_y$, for all side-chamber lengths at two representative Reynolds numbers ($\Re_{\mathrm{in}} = 50$ and 500) (Fig.~\ref{fig:Energy1}). Increasing $L_{S,j}$ has negligible effects on $\langle \overline{k_{\mathrm{str}}} \rangle$ over most of the main channel (Fig.~\ref{fig:Energy1}(a) and (d)), whereas $\langle \overline{k_{\mathrm{sec}}} \rangle$ decreases systematically with increasing $L_{S,j}$ (Fig.~\ref{fig:Energy1}(b) and (e)). 
Consistent with the recirculation observed in Fig.~\ref{fig:Case1_stream}, increasing $L_{S,j}$ results in additional recirculation within the side chamber and reduces the secondary-flow kinetic energy, consequently delivering substantially weaker secondary flow to the main channel. For $L_{S,j}/L_{S,\mathrm{ref}}=3$, $\langle k_{\mathrm{sec}}\rangle$ at mid-channel is approximately 93\% and 85\% lower than that of the reference case ($L_{S,j}/L_{S,\mathrm{ref}}=1$) at $\Re_{\mathrm{in}}=50$ and $500$, respectively.

\begin{figure}
\centerline{\includegraphics[scale=0.3]{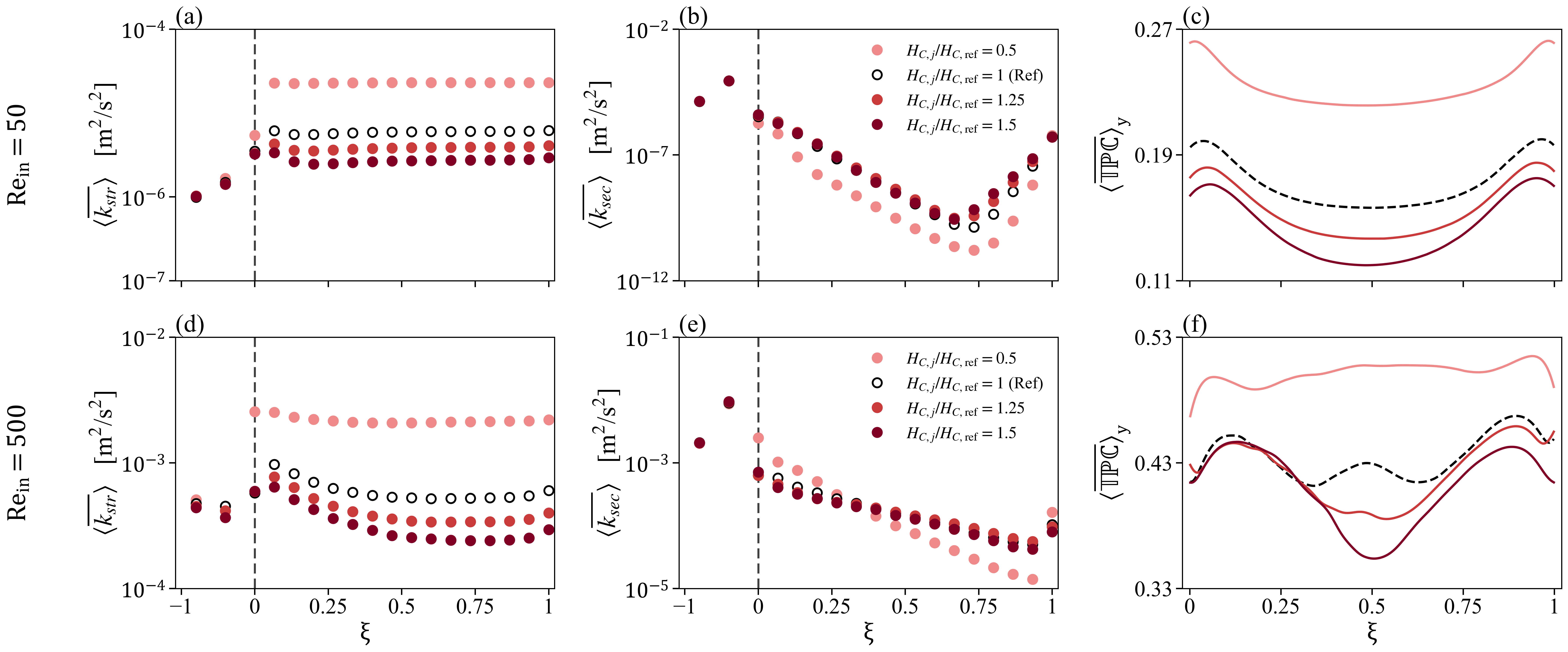}}
\caption{\label{fig:Energy2} Streamwise and secondary-flow kinetic energies and temperature polarization for varying main-channel height at $\Re_{\mathrm{in}}=50$ (top) and 500 (bottom), respectively. (a, d) time-spatial-averaged streamwise kinetic energy ($\langle \overline{k_{\mathrm{str}}}\rangle$)
(b, e) time-spatial-averaged secondary-flow kinetic energy, ($\langle \overline{k_{\mathrm{sec}}}\rangle$), and
(c, f) time-spanwise-averaged temperature polarization coefficient ($\langle \overline{\mathbb{TPC}}\rangle_y$
). Negative $\xi$ corresponds to the side chamber, and the dashed line denotes the chamber-to-channel junction at $\xi=0$.}
\end{figure}

\begin{figure}
\centerline{\includegraphics[scale=0.3]{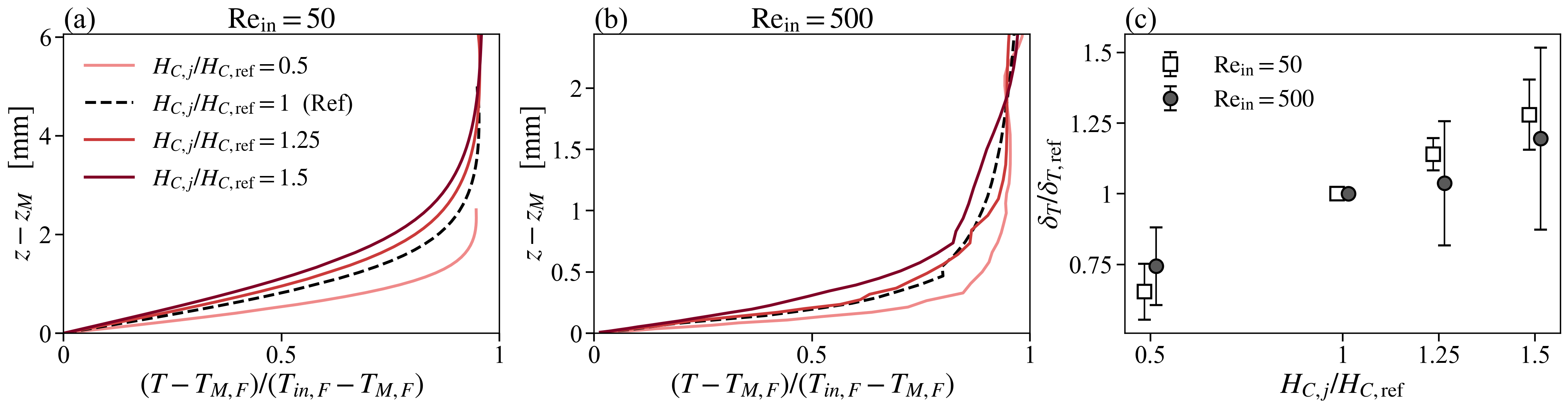}}
\caption{\label{fig:Case3_thermal} Feed-side temperature profiles normal to the membrane at mid-channel ($\xi=0.5$) for varying main-channel height at (a) $\Re_{\mathrm{in}}=50$ and (b) $\Re_{\mathrm{in}}=500$. (c) Normalized thermal-boundary-layer thickness ($\delta_T/\delta_{T,\mathrm{ref}}$) as a function of $H_{C,j}/H_{C,\mathrm{ref}}$ at both Reynolds numbers.}
\end{figure}

In contrast to side chamber lengths, increasing $H_{C,j}$ substantially decreases $\langle \overline{k_{\mathrm{str}}} \rangle$ at both Reynolds numbers (Figs.~\ref{fig:Energy2}(a) and (d)) due to a reduction in the main-channel velocity ($U_b$) given a fixed flow rate. Consistent with the generalized L\'ev\^eque scaling for the thermal boundary layer ($\delta_T\propto(D_h/U_b)^{1/3}$) \cite{Martin2002}, the increase in $D_h$ and decrease in $U_b$ are expected to increase the thermal-boundary-layer thickness. The temperature profiles at mid-channel ($\xi=0.5$) confirm this trend, with the thermal boundary layer becoming thicker with increasing $H_{C,j}/H_{C,\mathrm{ref}}$ at both Reynolds numbers (Figs.~\ref{fig:Case3_thermal}(a) and (b)). To quantify this trend, $\delta_T$ was defined as the wall-normal distance at which the normalized temperature reaches 0.99. The mean normalized boundary-layer thickness ($\delta_T/\delta_{T,\mathrm{ref}}$) increases with $H_{C,j}/H_{C,\mathrm{ref}}$ at both Reynolds numbers (Fig.~\ref{fig:Case3_thermal}(c)). As a result, the weaker near-membrane convective transport is consistent with the decrease in $\langle \overline{\mathbb{TPC}} \rangle_y$ (stronger TP) as $H_{C,j}$ increases. Changes in $H_{C,j}$ have relatively small effects on $\langle \overline{k_{\mathrm{sec}}} \rangle$, except for the smallest channel height ($H_{C,j}/H_{C,\mathrm{ref}}=0.5$), likely due to the formation of two pairs of Dean vortices. Overall, our results show that the main effect of $H_{C,j}$ on temperature polarization is associated with streamwise transport and thermal-boundary-layer development.

\section{\label{sec:discuss1}Discussion}
\subsection{Reynolds-number-dependent entrance effects on temperature polarization\label{subsec:discuss1}}

\begin{figure}
\centerline{\includegraphics[scale=0.25]{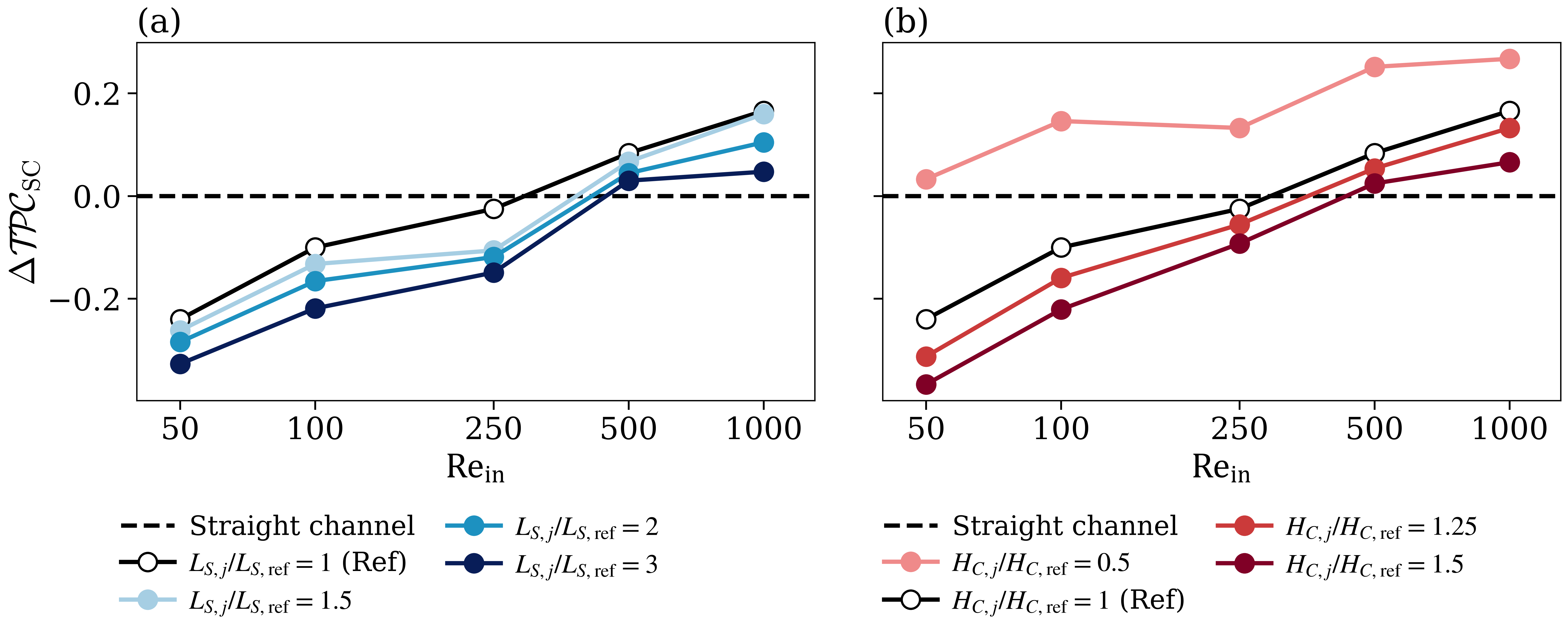}}
\caption{\label{fig:TPC_SC} Relative deviation of the time-spatial-averaged temperature polarization coefficient from the straight-channel baseline ($\Delta\mathcal{TPC}_{\mathrm{SC}}$) as a function of $\Re_{\mathrm{in}}$ for varying (a) side-chamber length and (b) main-channel height. Zero indicates agreement with the straight-channel baseline, while negative and positive values indicate stronger and weaker temperature polarization, respectively.}
\end{figure}

Fig.~\ref{fig:TPC_SC} shows the relative deviation of the time-spatial-averaged temperature polarization coefficient from the straight-channel baseline ($\Delta{\mathcal{TPC}_{\mathrm{SC}}} = (\langle \overline{\mathbb{TPC}} \rangle-\langle \overline{\mathbb{TPC}} \rangle_{\mathrm{SC}})/\langle \overline{\mathbb{TPC}} \rangle_{\mathrm{SC}}$). Negative and positive values indicate stronger and weaker TP than the straight-channel baseline, respectively. For most geometries, the deviation is negative at lower $\Re_{\mathrm{in}}$ and becomes positive as $\Re_{\mathrm{in}}$ increases, showing that the entrance-induced effect changes in both magnitude and direction with flow conditions. The case with $H_{C,j}/H_{C,\mathrm{ref}}=0.5$ is an exception and remains above the straight-channel baseline over the Reynolds-number range considered.

This Reynolds-number-dependent reversal reveals the balance among secondary-flow strength, downstream persistence, and streamwise transport. For all Reynolds numbers simulated, counter-rotating vortices form near the main-channel entrance, indicating that part of the streamwise motion is redirected into spanwise motion \cite{Haines2013}. At lower Reynolds numbers (e.g., $\Re_{\mathrm{in}}=50$), these vortices decay rapidly, and only weak spanwise variations in $\enstmean{\mathbb{TPC}}$ remain by mid-channel. The Dean vortices, therefore, do not persist long and strong enough to result in sustained spanwise heat transport over most of the membrane. In this regime, kinetic energy is redirected into short-lived secondary motion that decays before achieving a sustained mixing benefit. The reduction in streamwise transport is therefore not compensated by downstream secondary-flow transport, which is consistent with the stronger TP observed relative to the straight-channel baseline. At higher Reynolds numbers (e.g., $\Re_{\mathrm{in}}=500$), the vortices persist farther downstream, and their spatial structure is reflected in the TP distribution on the membrane. The secondary motion therefore remains effective over a larger fraction of the membrane, and the resulting secondary-flow transport can offset the reduction in streamwise transport associated with vortex formation. Thus, the entrance-induced effect depends not only on the formation of Dean vortices, but also on whether the resulting vortices persist downstream.

Regarding the impacts of geometry on temperature polarization, increasing $L_{S,j}$ reduces the secondary-flow kinetic energy ($\left\langle \overline{k_{\mathrm{sec}}}\right\rangle$) entering the main channel and decreases $\langle \overline{\mathbb{TPC}} \rangle$ (stronger TP) across the Reynolds numbers considered. At lower $\Re_{\mathrm{in}}$, the Dean vortices decay rapidly and result in only limited downstream secondary-flow transport. Stronger secondary flow associated with smaller $L_{S,j}$ can partially offset this limitation and brings the lab-scale module closer to the straight-channel baseline. Increasing $L_{S,j}$ further weakens the secondary flow, resulting in stronger TP and a larger deviation from the straight-channel baseline. At higher $\Re_{\mathrm{in}}$, the Dean vortices persist farther downstream, and the stronger secondary flow associated with smaller $L_{S,j}$ persists sufficiently downstream to weaken TP relative to the straight-channel baseline. Increasing $L_{S,j}$ weakens this secondary flow and therefore reduces the deviation from the straight-channel baseline. Thus, stronger secondary flow results in closer agreement with straight-channel behavior at lower Reynolds numbers, while weaker secondary flow results in closer agreement at higher Reynolds numbers.

Main-channel height affects temperature polarization through a different mechanism. Increasing $H_{C,j}$ lowers the bulk streamwise velocity and streamwise kinetic energy, which leads to a thicker thermal boundary layer and stronger TP. At lower Reynolds numbers, the Dean vortices decay rapidly and provide limited mixing downstream. The reduction in streamwise kinetic energy therefore further strengthens TP, resulting in a larger deviation from the straight-channel baseline as $H_{C,j}$ increases. At higher Reynolds numbers, the Dean vortices persist farther downstream and weaken TP relative to the straight-channel baseline. Increasing $H_{C,j}$ reduces the streamwise kinetic energy and strengthens TP, partially offsetting the effect of the persistent secondary flow and bringing the module closer to the straight-channel baseline.

Although $L_{S,j}$ and $H_{C,j}$ modify the flow through different mechanisms, both cases show that the geometric configuration closest to straight-channel behavior depends on the Reynolds number. At lower Reynolds numbers, the Dean vortices decay rapidly, so further reductions in secondary-flow strength or streamwise kinetic energy generally strengthen TP. At higher Reynolds numbers, the vortices persist farther downstream and weaken TP, so reducing the secondary-flow strength or streamwise kinetic energy can bring the lab-scale module closer to the straight-channel baseline. The resulting entrance effect, therefore, depends on the balance between secondary-flow strength, downstream persistence, and streamwise transport.

\subsection{\label{sec:recomm}Implications for lab-scale membrane characterization and heat-transfer correlations}

A common practice in MD studies is to use lab-scale modules to evaluate membrane performance and develop empirical heat-transfer correlations (e.g., Nusselt-Reynolds correlations) for larger-scale modeling \cite{Mohammad2016}. However, entrance-induced hydrodynamics in these modules can alter TP, introducing uncertainty into both applications. For membrane characterization, such variations can influence measured permeate flux and membrane-surface conditions relevant to scaling \cite{Christie2022-zd}, complicating comparisons across testing setups. For heat-transfer correlations, entrance-induced effects may become embedded in the empirical relationships, introducing geometry-specific contributions that are not representative of heat transfer in longer MD channels.

Our results further show that the magnitude and direction of these entrance-induced deviations depend on the Reynolds number,with stronger TP at lower Reynolds numbers and weaker TP at higher Reynolds numbers relative to the straight-channel baseline for most geometries. The geometric configuration that minimizes the deviation from the straight-channel baseline at lower Reynolds numbers may not minimize the deviation at higher Reynolds numbers. For most geometries in this study, there is a Reynolds number at which the average temperature polarization coefficient of the lab-scale module matches that of the entrance-minimized straight-channel baseline, and this Reynolds number varies with module geometry. The entrance-induced deviation, therefore, cannot be represented by a Reynolds-number-independent correction over the Reynolds-number range used to develop a heat-transfer correlation.

For membrane characterization, entrance-induced TP should be considered when interpreting permeate-flux and scaling measurements across different lab-scale designs, particularly when evaluating new membrane materials. Two lab-scale modules with the same membrane-channel dimensions and operating Reynolds number can exhibit different TP if their entrance geometries generate secondary flows with different strengths and downstream persistence. These geometry-specific differences also raise concerns for larger-scale modeling, as heat-transfer correlations derived from short lab-scale modules may not apply directly to other module configurations. The entrance geometry and Reynolds-number range should therefore be reported when comparing membrane performance or developing heat-transfer correlations from lab-scale measurements. Future investigations should determine whether similar Reynolds-number-dependent entrance effects occur in other module geometries and MD configurations, including vacuum, air-gap, sweeping-gas, and osmotic membrane distillation.

\section{Conclusion}

We investigated the impacts of side-chamber length and main-channel height on entrance-induced hydrodynamics and temperature polarization in three-dimensional, lab-scale DCMD modules over $\Re_{\mathrm{in}}=10$--$1000$. Counter-rotating streamwise vortices formed near the main-channel entrance across all geometries, but their downstream persistence increased significantly with $\Re_{\mathrm{in}}$. At lower $\Re_{\mathrm{in}}$, the vortices mostly decayed before reaching mid-channel and resulted in only weak spanwise variations in temperature polarization. At higher $\Re_{\mathrm{in}}$, more streamwise-vorticity magnitude persisted downstream, and the resulting vortex structures were reflected in the spatial distribution of temperature polarization. The formation of entrance-induced vortices alone is therefore insufficient to characterize their influence on membrane heat transfer. Their downstream persistence and spatial structure must also be considered.

Side-chamber length and main-channel height modified temperature polarization through different flow mechanisms. Increasing $L_{S,j}$ generated additional recirculation within the side chamber and reduced the secondary-flow kinetic energy delivered to the main channel. The average temperature polarization coefficient consequently decreased, indicating stronger temperature polarization, except at the lowest Reynolds number where the geometric effect was small. The persistence of streamwise vorticity varied with $L_{S,j}$ and $\Re_{\mathrm{in}}$, showing that secondary-flow strength and downstream persistence do not necessarily change in the same manner. In contrast, increasing $H_{C,j}$ reduced the bulk velocity and streamwise kinetic energy. The thermal boundary layer became thicker as $H_{C,j}$ increased, which led to stronger temperature polarization. Thus, the geometric dependence of temperature polarization cannot be described by secondary-flow strength alone, but also reflects changes in downstream vortex evolution, streamwise transport, and thermal-boundary-layer development.

When compared with the entrance-minimized straight-channel baseline, most geometries resulted in stronger temperature polarization at lower $\Re_{\mathrm{in}}$ and weaker temperature polarization at higher $\Re_{\mathrm{in}}$. The entrance-induced deviation therefore changes not only in magnitude but also in direction as the flow condition changes. As a result, a geometric modification that reduces the deviation from entrance-minimized straight-channel behavior at one Reynolds number is unlikely to reduce it at another. This Reynolds-number dependence limits the use of a single correction for entrance effects in short lab-scale modules. Empirical heat-transfer correlations obtained from these modules can contain entrance-specific contributions that depend on both geometry and operating condition. The entrance geometry and Reynolds-number range should therefore be reported and considered when such correlations are developed, compared, or applied beyond the configuration from which they were obtained. Accounting for these effects can also improve the interpretation of membrane performance measurements obtained from different lab-scale modules.

\section*{Acknowledgements}

Portions of this research were conducted with the advanced computing resources provided by Texas A\&M High Performance Research Computing.

\appendix

\section{Grid convergence study}
\label{sec:convergence}
To determine the grid resolution needed for accurate simulation, we performed grid convergence studies with the reference case described in \Cref{sec:geo-parmeters}. The simulations were set up as described in \Cref{sec:setup} using five grid resolutions ($\Delta x =  5\times 10^{-4}, 3.75\times 10^{-4}, 2.5\times 10^{-4}, 1.875\times 10^{-4},$ and $1.25 \times 10^{-4}\,\si{m}$). To evaluate convergence, we define both the time–$y$-averaged temperature and the time–$x$-averaged temperature in the feedwater, respectively, as

\begin{subequations}
\begin{equation}
        \langle \overline{T_{M,F}} \rangle_{y,t} (x) = \frac{1}{N_y N_t} \sum_{j=1}^{N_y} \sum_{n=1}^{N_t} T_{M,F} (x, y_j, t_n). 
\end{equation}

\begin{equation}
        \langle \overline{T_{M,F}} \rangle_{x,t} (y) = \frac{1}{N_x N_t} \sum_{i=1}^{N_x} \sum_{n=1}^{N_t} T_{M,F} (x_i, y, t_n), 
\end{equation}
\end{subequations}

\begin{figure}[H]
    \centerline{
     {\includegraphics[scale=0.8]{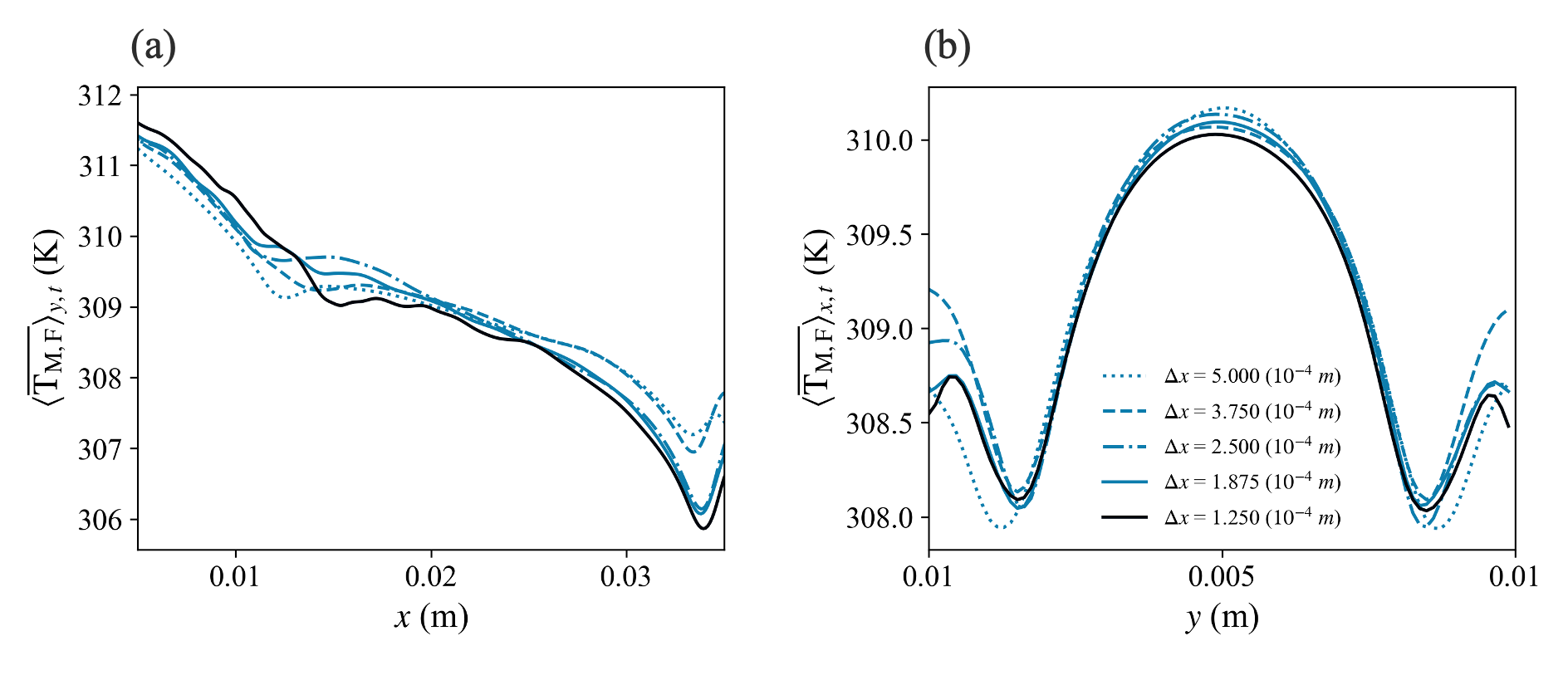}}}
     \caption{Grid-convergence analysis showing (a) time-$y$-averaged feedwater temperature as a function of $x$ and (b) time-$x$-averaged feedwater temperature as a function of $y$ for $\Delta x=5\times10^{-4}$, $3.75\times10^{-4}$, $2.5\times10^{-4}$, $1.875\times10^{-4}$, and $1.25\times10^{-4}$ m.}
     \label{fig:grid_conv}
\end{figure}

\noindent Figs.~\ref{fig:grid_conv} (a) and (b) show $\langle \overline{T_{M,F}} \rangle_{y,t}$ and $\langle \overline{T_{M,F}} \rangle_{x,t}$ as functions of $x$ and $y$, respectively, with color indicating different grid resolution ($\Delta x$). Overall, we observed convergence in both $\langle \overline{T_{M,F}} \rangle_{y,t}$ and $\langle \overline{T_{M,F}} \rangle_{x,t}$ as $\Delta x$ decreases. Based on these results, we select a grid resolution of $\Delta{x} = \SI{0.125e-3}{\meter}$ for all subsequent simulations.
 
\section{Numerical verification and validation}
\label{sec:validation}

\begin{figure}[H]
    \centerline{
     {\includegraphics[scale=0.5]{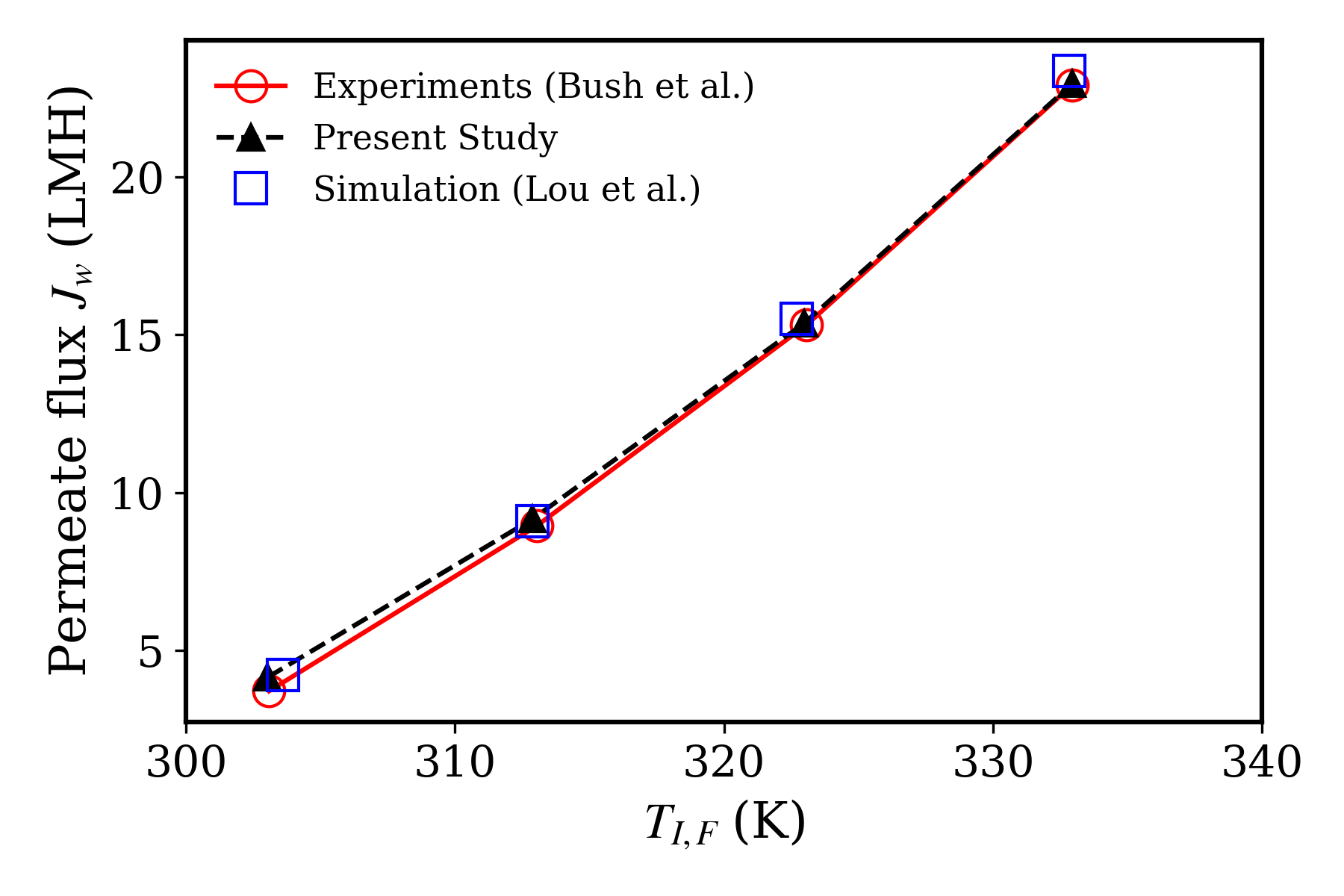}}}
     \caption{Model verification and validation based on permeate flux ($J_w$) as a function of feed-side inlet temperature ($T_{I,F}$), with the permeate-side inlet temperature fixed at $T_{I,P}=293.15$~K (20~\unit{\celsius}). The present simulation results are compared with the experimental measurements by Bush \emph{et al.} \cite{Bush2016} and numerical results by Lou \emph{et al.} \cite{Lou2019}.}
     \label{fig:validation}
\end{figure}

To verify and validate the solver used in this study, we performed two-dimensional DCMD simulations following Lou \emph{et al.} \cite{Lou2019} and our previous study \cite{Yao2023}. We consider a two-dimensional DCMD module consisting of a feed channel and a permeate channel with lengths of $L=177.8 \ \si{mm}$ and heights of $H=3.175 \ \si{mm}$, respectively. The experimental configuration was reported by Lou \emph{et al.} \cite{Lou2019}, with additional details of the experimental setup provided by Bush \emph{et al.} \cite{Bush2016}. The experiments used a polypropylene membrane with a nominal pore size of $0.2~\si{\micro\meter}$. The feed-side inlet temperature $T_{I,F}$ was varied from 303.15 to 333.15~K (30-60~\unit{\celsius}), while the permeate-side inlet temperature $T_{I,P}$ was kept at 293.15~K (20~\unit{\celsius}). Fig.~\ref{fig:validation} shows the permeate flux $J_w$ as a function of $T_{I,F}$. Using the calibrated permeability, the predicted permeate flux agrees closely with the experimental measurements and numerical results reported by Lou \emph{et al.} \cite{Lou2019}.

\section{Effects of side-chamber height}
\label{app:Hs}

\begin{figure}[H]
    \centerline{
     {\includegraphics[scale=0.3]{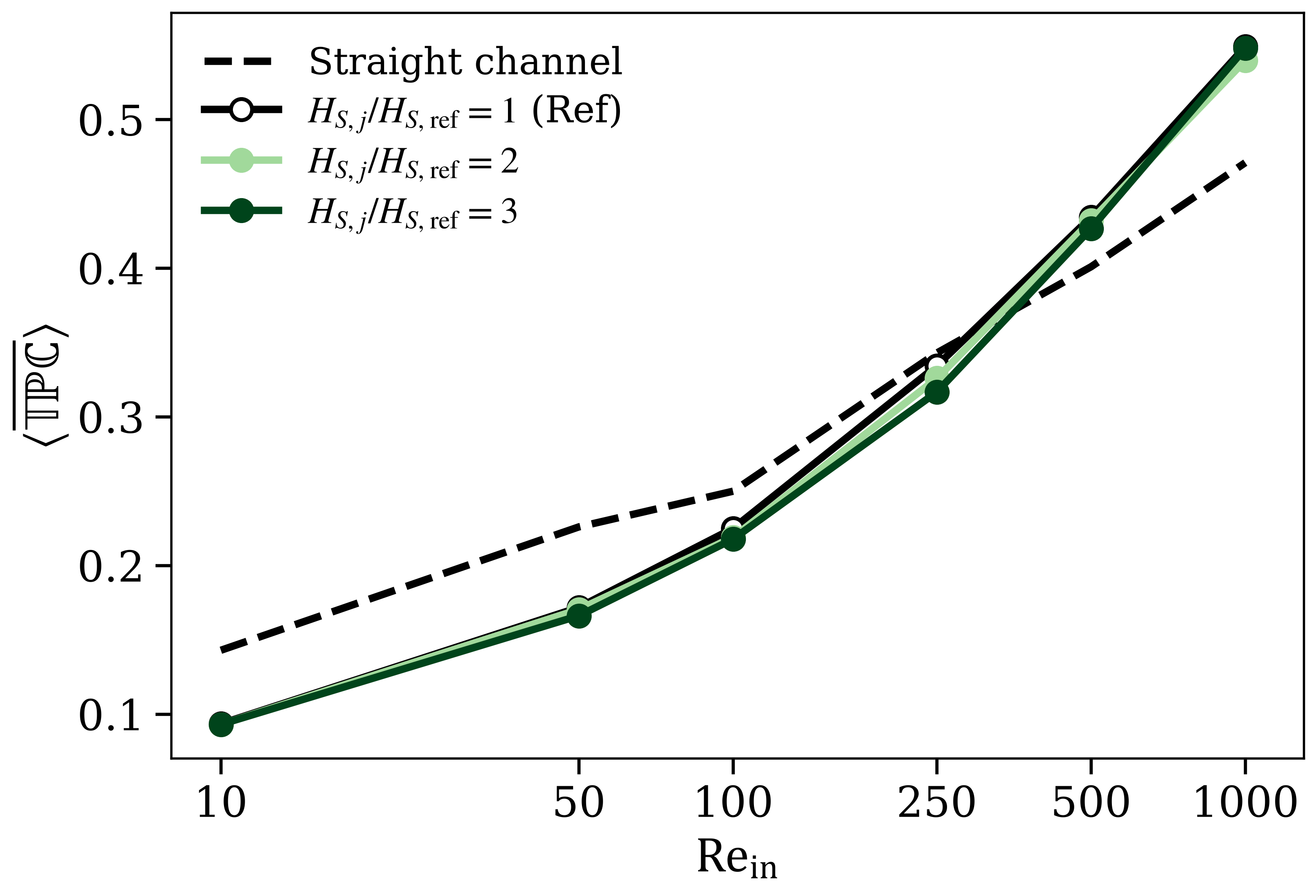}}}
     \caption{Time-spatial-averaged temperature polarization coefficient ($\enstmean{\mathbb{TPC}}$) as a function of $\Re_{\mathrm{in}}$ for varying side-chamber height. The dashed line refers to the ideal straight-channel baseline.}
     \label{fig:appendix_TPC_HS}
\end{figure}

The effects of side-chamber height were evaluated using
$H_{S,j}/H_{S,\mathrm{ref}}=\{1,2,3\}$, with
$H_{S,\mathrm{ref}}=10$~\unit{\mm}.
As shown in Fig.~\ref{fig:appendix_TPC_HS}, variations in $H_{S,j}$ result in only minor differences in the time-spatial-averaged temperature polarization coefficient ($\enstmean{\mathbb{TPC}}$) over the Reynolds-number range considered. These differences are insignificant as compared to those associated with side-chamber length and main-channel height discussed in the main text. Side-chamber height is therefore not considered further in the main analysis.

\section{Temperature polarization and vorticity fields for all geometries}

\begin{figure}[H]
    \centerline{
     {\includegraphics[scale=0.5]{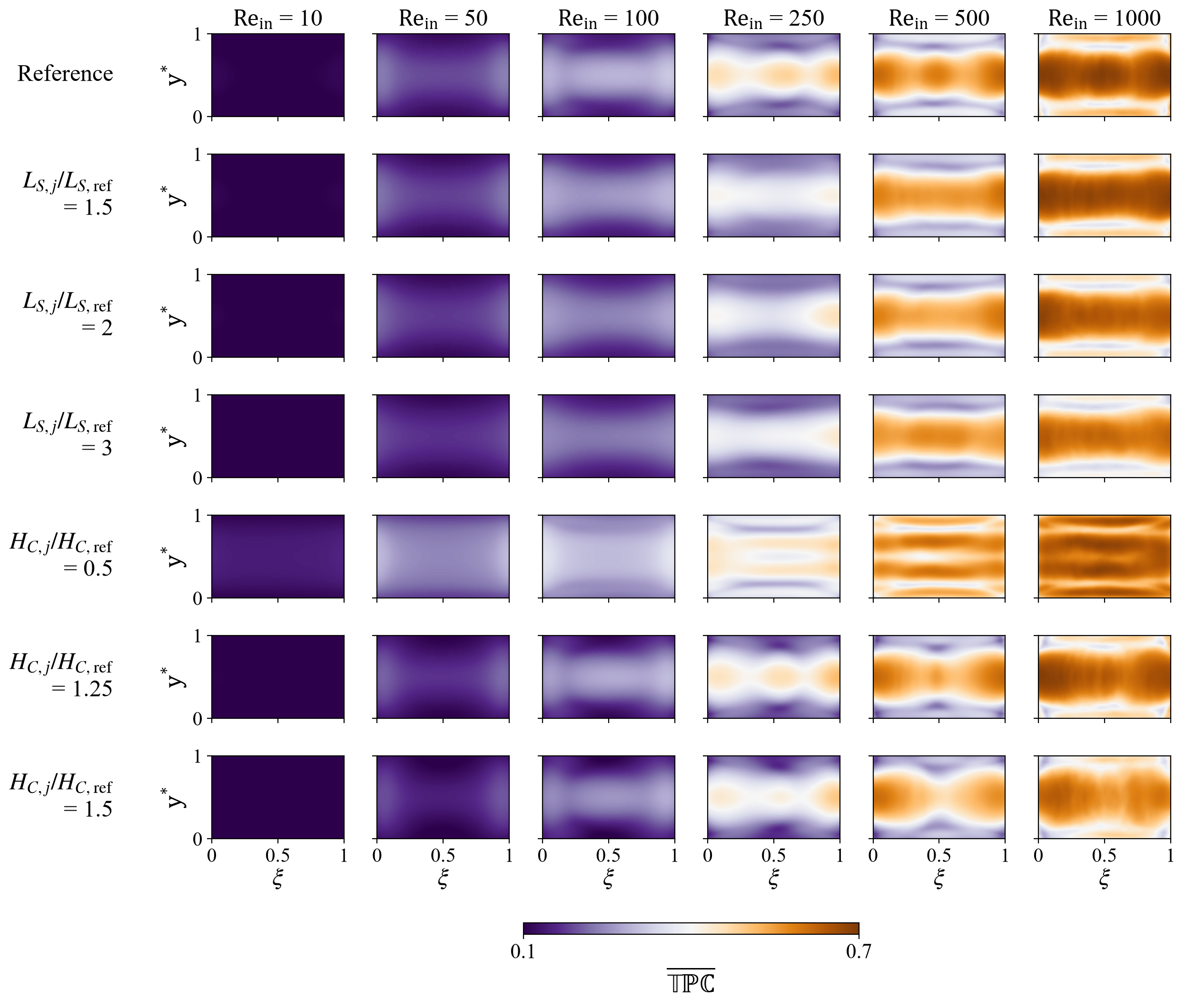}}}
     \caption{Time-averaged temperature polarization coefficient ($\tmean{\mathbb{TPC}}$) on the membrane surface for all simulated geometries and inlet Reynolds numbers. Rows correspond to module geometries and columns to $\Re_{\mathrm{in}}$}
     \label{fig:appendix_TPC_mem}
\end{figure}

\begin{figure}[H]
    \centerline{
     {\includegraphics[scale=0.5]{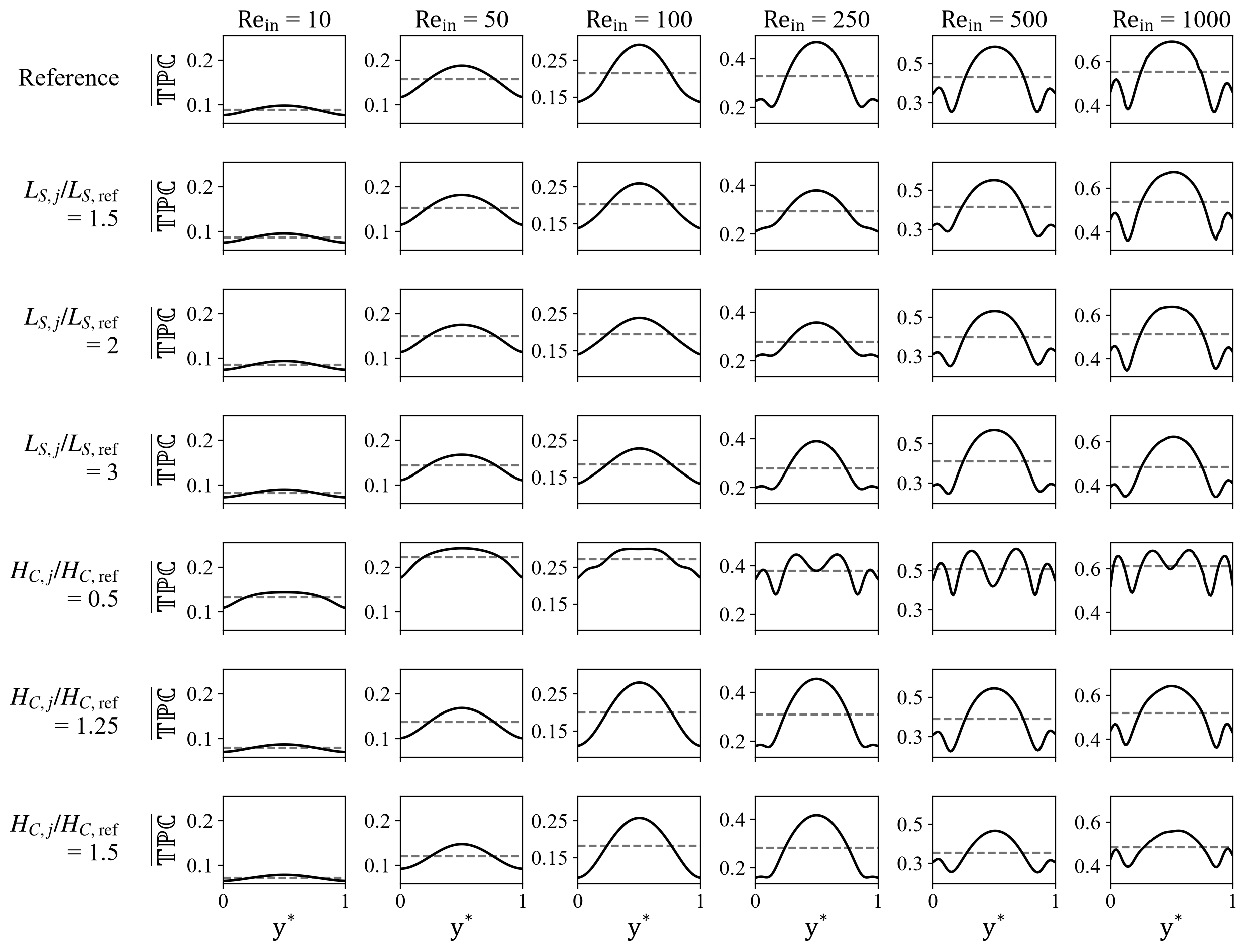}}}
     \caption{Spanwise profiles of the time-averaged temperature polarization coefficient ($\tmean{\mathbb{TPC}}$) at mid-channel ($\xi=0.5$) for all simulated geometries and inlet Reynolds numbers. Rows correspond to module geometries and columns to $\Re_{\mathrm{in}}$.}
     \label{fig:appendix_TPC_center}
\end{figure}

\begin{figure}[H]
    \centerline{
     {\includegraphics[scale=0.25]{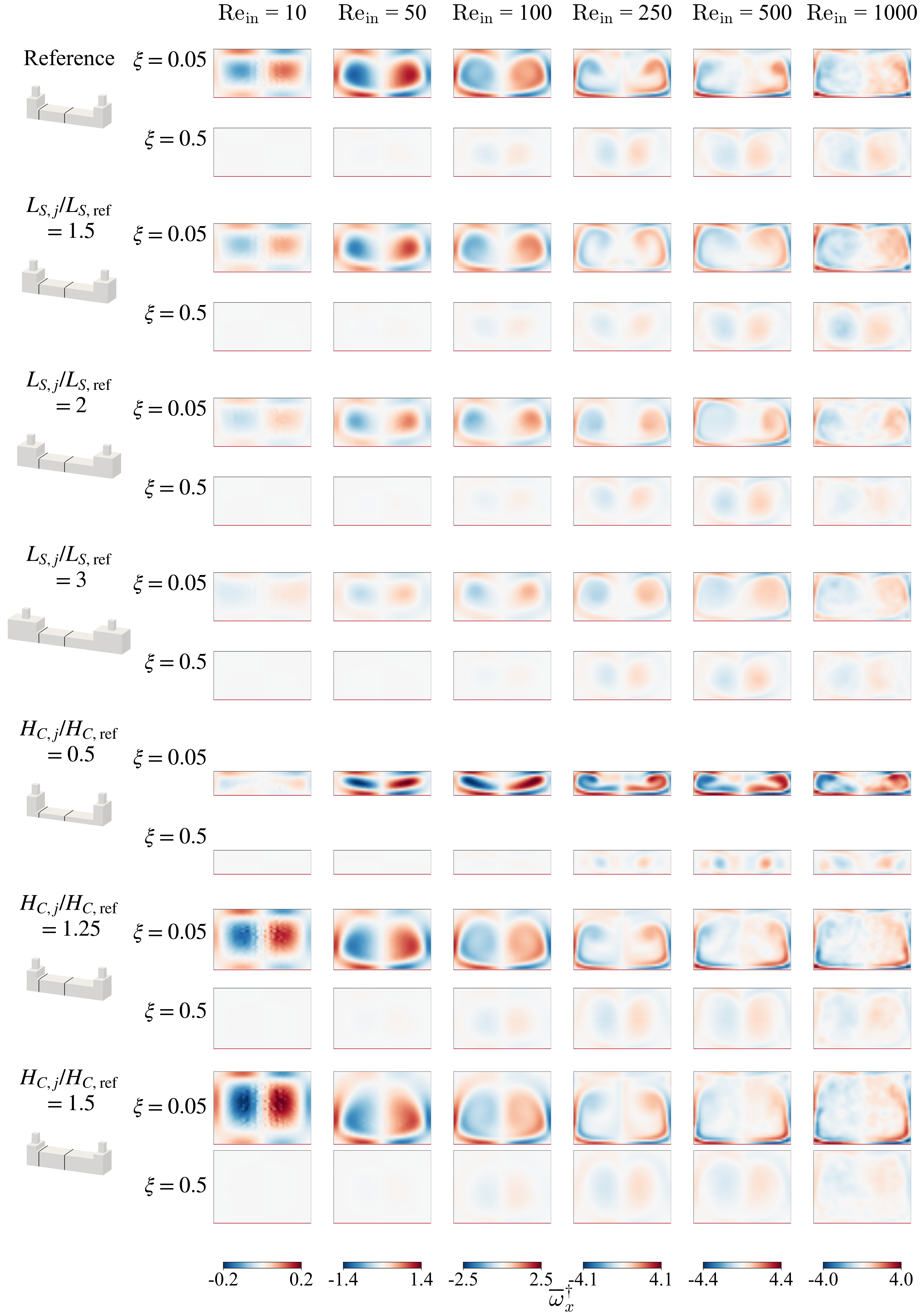}}}
     \caption{Normalized time-averaged streamwise vorticity ($\overline{\omega_x^\dagger}$) at the upstream ($\xi=0.05$) and mid-channel ($\xi=0.5$) cross sections for all simulated geometries and inlet Reynolds numbers. Rows correspond to module geometries and streamwise locations, while columns correspond to $\Re_{\mathrm{in}}$.}
     \label{fig:appendix_TPC_vort}
\end{figure}

The complete temperature-polarization and streamwise-vorticity results for all simulated geometries and Reynolds numbers are provided in Figs.~\ref{fig:appendix_TPC_mem}-\ref{fig:appendix_TPC_vort}.

\bibliographystyle{elsarticle-num}

\end{document}